\newif\ifacm
\acmtrue

\ifacm
\documentclass[acmsmall]{acmart}
\else
\documentclass[11pt]{article}
\usepackage[margin=1in]{geometry}
\fi

\usepackage{amsmath}
\usepackage{amssymb}
\usepackage{mathpartir}
\usepackage{setspace}
\usepackage{xspace}
\usepackage{alltt}
\usepackage{prooftrees}
\usepackage{todonotes}
\usepackage{xcolor}
\usepackage{tikz}
\usepackage{listings}
\usepackage{syntax}
\usepackage{paralist}
\usepackage{rotating}
\usepackage{stmaryrd}
\usepackage{alltt}
\usepackage{moresize}
\usepackage{wrapfig}
\usepackage{lscape}
\usepackage{microtype}
\usepackage[skip=0pt]{caption}
\usepackage{url}

\usetikzlibrary{shapes}
\usetikzlibrary{automata}
\tikzstyle{circ}=[draw,circle,minimum size=0.4cm]

\tikzset{
  =>stealth',
  prop/.style={
    rectangle,
    rounded corners,
    draw=black, very thick,
    text centered
  },
  feature/.style={
    rectangle,
    draw=violet, very thick,
    text centered
  },
  conflict/.style={
    rectangle,
    draw=red,
    minimum width=1em,
    minimum height=1em
  },
  depend/.style={
    ->,
    draw=black, thick
  },
  confmember/.style={
    draw=red, very thick
  },
  mitigate/.style={
    ->,
    draw=blue, very thick
  },
}

\usepackage{supertabular}

\newcommand{\nlsays}[1]{\todo[color=green!40]{NL: #1}}
\newcommand{\pcsays}[1]{\todo[color=blue!40]{PC: #1}}
\newcommand{\arsays}[1]{\todo[color=purple!40]{AR: #1}}
\newcommand{\agsays}[1]{\todo[color=yellow!40]{AG: #1}}
\newcommand{\TODO}[1]{\todo[color=red!20]{TODO: #1}}
\newcommand{\ignore}[1]{ }

\DeclareMathOperator{\vis}{\mathsf{vis}}

\newcommand{\defn}{:=}

\newcommand{\Add}[1]{\mathsf{Add} \; #1}
\newcommand{\Sub}[1]{\mathsf{Sub} \; #1}

\newcommand{\guarded}[2]{ #1 \triangleright #2 }

\DeclareMathOperator{\eval}{\texttt{eval}}

\newcommand{\WCP}{\texttt{WCP}\xspace}

\newcommand{\eff}[1]{\textsf{eff} \; #1}

\newcommand{\CARD}{CARD\xspace}
\newcommand{\CARDs}{CARDs\xspace}

\newcommand{\fst}[1]{\mathsf{First}}

\newcommand{\app}{\mathtt{app}}

\newcommand{\id}{\mathsf{id}}

\newcommand{\denote}[1]{\llbracket #1 \rrbracket}

\newcommand{\ar}{\mathsf{ar}}

\renewcommand{\eff}{\mathsf{eff}}
\newcommand{\gc}{\mathsf{gc}}

\newcommand{\grd}{\mathsf{grd}}
\newcommand{\rval}{\mathsf{rval}}
\renewcommand{\eval}{\mathsf{eval}}
\newcommand{\op}{\mathsf{op}}

\newcommand{\ottdrule}[4][]{{\displaystyle\frac{\begin{array}{l}#2\end{array}}{#3}\quad\ottdrulename{#4}}}
\newcommand{\ottusedrule}[1]{\[#1\]}
\newcommand{\ottpremise}[1]{ #1 \\}
\newenvironment{ottdefnblock}[3][]{ \framebox{\mbox{#2}} \quad #3 \\[0pt]}{}

\newcommand{\ottnt}[1]{\mathit{#1}}

\newcommand{\ottkw}[1]{\mathbf{#1}}
\newcommand{\ottsym}[1]{#1}

\newcommand{\ottdrulename}[1]{\textsc{#1}}

\renewcommand{\ottpremise}[1]
  {#1 \qquad}
\renewcommand{\ottusedrule}[1]
  {\hfill\(#1\)\hfill}

\newcommand{\ottdruletypeXXvar}[1]{\ottdrule[#1]{%
}{
\Gamma  \ottsym{,}  \mathit{x}  \ottsym{:}  \tau  \vdash  \mathit{x}  \ottsym{:}  \tau}{%
{\ottdrulename{type\_var}}{}%
}}

\newcommand{\ottdruletypeXXlambda}[1]{\ottdrule[#1]{%
\ottpremise{\Gamma  \ottsym{,}  \mathit{x}  \ottsym{:}  \tau_{{\mathrm{1}}}  \vdash  \ottnt{t}  \ottsym{:}  \tau_{{\mathrm{2}}}}%
}{
\Gamma  \vdash  \lambda  \mathit{x}  \ottsym{.}  \ottnt{t}  \ottsym{:}  \tau_{{\mathrm{1}}}  \rightarrow  \tau_{{\mathrm{2}}}}{%
{\ottdrulename{type\_lambda}}{}%
}}

\newcommand{\ottdruletypeXXite}[1]{\ottdrule[#1]{%
\ottpremise{\Gamma  \vdash  \ottnt{t_{{\mathrm{1}}}}  \ottsym{:}   \mathtt{Bool} }%
\ottpremise{\Gamma  \ottsym{,}  \ottnt{t_{{\mathrm{1}}}}  \vdash  \ottnt{t_{{\mathrm{2}}}}  \ottsym{:}  \tau}%
\ottpremise{\Gamma  \ottsym{,}  \neg \, \ottnt{t_{{\mathrm{1}}}}  \vdash  \ottnt{t_{{\mathrm{3}}}}  \ottsym{:}  \tau}%
}{
\Gamma  \vdash  \ottkw{if} \, \ottnt{t_{{\mathrm{1}}}} \, \ottkw{then} \, \ottnt{t_{{\mathrm{2}}}} \, \ottkw{else} \, \ottnt{t_{{\mathrm{3}}}}  \ottsym{:}  \tau}{%
{\ottdrulename{type\_ite}}{}%
}}

\newcommand{\ottdruletypeXXq}[1]{\ottdrule[#1]{%
\ottpremise{\bullet  \vdash  \ottnt{c}  \ottsym{:}  \ottnt{C}}%
\ottpremise{\Gamma  \ottsym{,}  \mathit{x}  \ottsym{:}  \ottsym{\{}  \nu  \ottsym{:}  \ottnt{S}  \; | \;  \ottsym{[}  \ottnt{s}  \ottsym{/}  s_g  \ottsym{]}  \ottsym{[}  \nu  \ottsym{/}  s_r  \ottsym{]}  \llbracket  \ottnt{c}  \rrbracket  \ottsym{\}}  \vdash  \ottnt{t}  \ottsym{:}  \text{Op} \, \ottsym{(}  \ottsym{(S,E,C)}  \ottsym{,}  \ottnt{A}  \ottsym{,}  \varphi  \ottsym{)}}%
}{
\Gamma  \vdash  \emph{Q} \, \ottnt{c}  \triangleright  \mathit{x}  \ottsym{.}  \ottnt{t}  \ottsym{:}  \text{Op} \, \ottsym{(}  \ottsym{(S,E,C)}  \ottsym{,}  \ottnt{A}  \ottsym{,}  \varphi  \ottsym{)}}{%
{\ottdrulename{type\_q}}{}%
}}

\newcommand{\ottdruletypeXXr}[1]{\ottdrule[#1]{%
\ottpremise{\Gamma  \vdash  t_e  \ottsym{:}  \ottnt{E}}%
\ottpremise{\Gamma  \vdash  t_a  \ottsym{:}  \ottsym{\{}  \nu  \ottsym{:}  \ottnt{A}  \; | \;  \ottnt{s'}  \ottsym{=}  \llbracket  t_e  \rrbracket \, \ottsym{(}  \ottnt{s}  \ottsym{)}  \Rightarrow  \varphi  \ottsym{\}}}%
}{
\Gamma  \vdash  \emph{R} \, \ottsym{.}  \ottsym{(}  t_e  \ottsym{,}  t_a  \ottsym{)}  \ottsym{:}  \text{Op} \, \ottsym{(}  \ottsym{(S,E,C)}  \ottsym{,}  \ottnt{A}  \ottsym{,}  \varphi  \ottsym{)}}{%
{\ottdrulename{type\_r}}{}%
}}

\newcommand{\ottdrulequery}[1]{\ottdrule[#1]{%
\ottpremise{\ottsym{(}  \ottnt{s}  \ottsym{,}  s_x  \ottsym{)}  \models  \ottnt{c}}%
\ottpremise{\ottsym{[}  s_x  \ottsym{/}  \mathit{x}  \ottsym{]}  \ottnt{t} \, \Downarrow_{\lambda} \, \ottnt{t'}}%
}{
\ottsym{(}  \ottnt{s}  \ottsym{,}  \psi  \ottsym{,}  \emph{Q} \, \ottnt{c}  \triangleright  \mathit{x}  \ottsym{.}  \ottnt{t}  \ottsym{)}  \longmapsto  \ottsym{(}  \ottnt{s}  \ottsym{,}  \psi  \land  \ottsym{[}  s_x  \ottsym{/}  s_r  \ottsym{]}  \ottnt{c}  \ottsym{,}  \ottnt{t'}  \ottsym{)}}{%
{\ottdrulename{query}}{}%
}}

\newcommand{\ottdruledrift}[1]{\ottdrule[#1]{%
\ottpremise{\exists \, \ottnt{e}  \ottsym{:}  \ottnt{E}  .\;  \ottnt{s'}  \ottsym{=}  \llbracket  \ottnt{e}  \rrbracket \, \ottsym{(}  \ottnt{s}  \ottsym{)}}%
\ottpremise{\ottnt{s'}  \models  \psi}%
}{
\ottsym{(}  \ottnt{s}  \ottsym{,}  \psi  \ottsym{,}  \ottnt{t}  \ottsym{)}  \longmapsto  \ottsym{(}  \ottnt{s'}  \ottsym{,}  \psi  \ottsym{,}  \ottnt{t}  \ottsym{)}}{%
{\ottdrulename{drift}}{}%
}}

\newcommand{\ottdefnprocXXstep}[1]{\begin{ottdefnblock}[#1]{$\ottsym{(}  \ottnt{s}  \ottsym{,}  \psi  \ottsym{,}  \ottnt{t}  \ottsym{)}  \longmapsto  \ottsym{(}  \ottnt{s'}  \ottsym{,}  \psi'  \ottsym{,}  \ottnt{t'}  \ottsym{)}$}{}
\ottusedrule{\ottdrulequery{}}
\ottusedrule{\ottdruledrift{}}
\end{ottdefnblock}}

\newif\iffinal
\finaltrue

\iffinal
\renewcommand{\nlsays}[1]{ }
\renewcommand{\pcsays}[1]{ }
\renewcommand{\arsays}[1]{ }
\renewcommand{\agsays}[1]{ }
\renewcommand{\TODO}[1]{ }

\newcommand{\sectionnewpage}{}

\else

\renewcommand{\nlsays}[1]{\todo[inline,color=green!40]{NL: #1}}
\renewcommand{\pcsays}[1]{\todo[inline,color=blue!40]{PC: #1}}
\renewcommand{\arsays}[1]{\todo[inline,color=purple!40]{AR: #1}}
\renewcommand{\agsays}[1]{\todo[inline,color=yellow!40]{AG: #1}}
\renewcommand{\TODO}[1]{\todo[inline,color=red!20]{TODO: #1}}

\newcommand{\sectionnewpage}{\newpage}

\fi

\ifacm
\acmYear{2018}
\acmMonth{7}
\startPage{1}
\else
\usepackage{amsthm}
\theoremstyle{definition}
\newtheorem{theorem}{Theorem}[section]
\newtheorem{example}[theorem]{Example}
\newtheorem{definition}[theorem]{Definition}
\newtheorem{lemma}[theorem]{Lemma}
\fi

\begin{document}

\title{Conflict-Aware Replicated Data Types}

\ifacm

\author{Nicholas V. Lewchenko}
\affiliation{\institution{University of Colorado Boulder}}
\email{nicholas.lewchenko@colorado.edu}
\author{Arjun Radhakrishna}
\affiliation{\institution{Microsoft}}
\email{arradha@microsoft.com}
\author{Akash Gaonkar}
\affiliation{\institution{University of Colorado Boulder}}
\email{akash.gaonkar@colorado.edu}
\author{Pavol {\v C}ern\'y}
\affiliation{\institution{University of Colorado Boulder}}
\email{pavol.cerny@colorado.edu}

\setcopyright{none}
\settopmatter{printacmref=false}

\renewcommand\footnotetextcopyrightpermission[1]{} 

\else

\author[1]{Nicholas V. Lewchenko}
\author[2]{Arjun Radhakrishna}
\author[1]{Pavol {\v C}ern\'y}
\affil[1]{University of Colorado Boulder}
\affil[2]{Microsoft}
\date{}

\fi

\ifacm
\else
\maketitle
\fi

\begin{abstract}
  {\bf Abstract}
  We introduce Conflict-Aware Replicated Data Types (CARDs).
  CARDs are significantly more expressive than Conflict-free Replicated Data
  Types (CRDTs) as they support operations that can conflict with each other.
  Introducing conflicting operations typically brings the need to block an
  operation in at least some executions, leading to difficulties in programming
  and reasoning about correctness, as well as potential inefficiencies in
  implementation.
  \hspace{\parindent}The salient aspect of CARDs is that they allow ease
  of programming and reasoning about programs comparable to CRDTs, while
  enabling algorithmic inference of conflicts so that an operation is
  blocked only when necessary.
  The key idea is to have a language that allows associating with each operation a two-state predicate called
  a {\em consistency guard} that relates the state of the replica on which the
  operation is executing to a global state (which is never computed).
  The consistency guards bring three advantages.
  First, a programmer developing an operation needs only to choose a consistency
  guard that states what the operation will rely on. In particular, they do not
  need to consider the operation conflicts with other operation. This allows
  purely {\em modular reasoning}.
  Second, we show that consistency guard allow reducing the complexity of
  reasoning needed to prove invariants that hold as CARD operations are
  executing. The reason is that consistency guard allow reducing the reasoning
  about concurrency among operations to purely {\em sequential reasoning}.
  Third, conflicts among operations can be algorithmically inferred by
  checking whether the effect of one operation preserves the consistency guard
  of another operation.
  We substantiate these claims by introducing a language for writing CARD
  operations.
  The language is dependently typed, and the type checking rules are
  based on the modular and sequential reasoning allowed by consistency guards.
  We also show how conflicts can be inferred at compile time, and the resulting
  constraints on executions can be enforced at runtime.
  We
  empirically show that the inference needed to detect conflicts between
  operations is well within the scope of
  current SMT solvers.
\end{abstract}

\ifacm
\maketitle
\else
\fi

\section{Introduction}
\label{sec:intro}

Conflict-free replicated data types (CRDTs) have quickly gained traction in
large-scale distributed
systems~\cite{CRDTs,Attiya:2016,Mehdi:2014,Nedelec:2013,GoogleDocs,IPFSCRDTs,Brown:2014}.
They allow operations to execute efficiently and independently across different
replicas without coordination while still guaranteeing strong eventual
consistency.
CRDTs rely on the fact that their operations are
\emph{conflict-free} (commutable).
However, the assumption of conflict-freedom is broken in many practical
scenarios
  either due to the presence of inherently conflicting operations, or
  due to the need for maintaining invariants on the data structure.

There have been several attempts to add conflicting operations to CRDTs using
mixed-consistency and tunable-consistency extensions in both academia and
industry~\cite{Gotsman:2016,Balegas:2015,Lakshman:2010,ECDS,RedBlue,Li:2014}.
However, most of these systems suffer from one of several drawbacks:
\begin{inparaenum}[(i)]
\item The programmer has to explicitly reason about and state conflicts for each
pair of operations~\cite{Gotsman:2016} or choose a consistency level (sequential
or eventual consistency in~\cite{RedBlue}) for each operation. These tasks
cannot be done modularly, that is, separately for each operation.
\item The programmer can specify consistency for each operation in isolation,
but the overall consistency model does not give clear guarantees. For
example, in Cassandra~\cite{Lakshman:2010}, a programmer can specify that an
operation can execute with coordination across just a small number of replicas.
However, it is not clear what consistency guarantees this provides the user.
\end{inparaenum}

\paragraph{Conflict-Aware Replicated Data-Types.} We present a novel extension
of CRDTs dubbed \emph{conflict-aware replicated data-types} (CARDs), which
support operations that might not be conflict-free.

From the perspective of the user of CARD operations, CARDs guarantee:
\begin{inparaenum}
\item[(a)] {\em strong eventual consistency} (SEC): all the replicas should
eventually process the (emitted effects of) operations and should agree on the
final value~\cite{CRDTs},
\item[(b)] {\em availability}: replicas should operate without
blocking coordination whenever possible---that is, whenever the operations do
not conflict, and
\item[(c)] preservation of {\em application-specific invariants}.
\end{inparaenum}

From the perspective of the developer of CARD operations, CARDs guarantee:
\begin{inparaenum}
\item[(d)] modular consistency specifications where the assumptions that an
operation relies on are stated with only that operation in mind, and allow
purely {\em modular reasoning}
\item[(e)] proof system where the reasoning about concurrent behavior is reduced
to {\em sequential reasoning}, and
\item[(f)] {\em automated detection of conflicts} between operations.
\end{inparaenum}

\paragraph{Execution model} CARD operations are executed by a network of
replicas. A client can ask a replica to execute an operation. The replica
evaluates the operation, provides a return value to the client, and sends the
{\em effect} of the operation to all the other replicas. The effect is a state transformer (for instance, it does not compute the return value) that the other
replicas use to update their states.

\paragraph{Consistency guards} The key idea of our approach is to introduce a
programming language that allows specifying consistency requirements for each
operation separately. The consistency requirements for an operation are
specified using a two-state predicate called a {\em consistency guard}. The
guard relates the replica state and the global state. An operation can rely on a
guard while it (the operation) is executing to ensure that no operations that
could break the guard are run in parallel.

\begin{figure}
  \begin{tikzpicture}[font=\tiny]
  \hspace*{-4mm}%
  \tikzset{cpt/.style={draw=black,fill=white,circle,inner sep=.5mm}}
  \def\repgap{1cm}

  \begin{scope}[xshift=-1cm]
    \coordinate[cpt,draw=white] (origin);
    \node[cpt,purple,label=left:{$\emptyset$}] (r1s) [left=\repgap of origin] {};
    \node[cpt,purple,label=left:{$\{(3, 5)\}$}] (r1m) [below=5mm of r1s] {};

    \node[cpt,blue,label=right:{$\{(3, 5)\}$}] (r2s) [below right=2mm and \repgap of r1m] {};
    \node[cpt,blue,label=right:{$\{(3, 5),(3, 7)\}$}] (r2m) [below=5mm of r2s] {};

    \node[cpt,label=left:{$\{(3, 5),(3, 7)\}$}] (r1e) [below left=2mm and \repgap of r2m] {};

    \path (origin) -- node [below=2.4cm] {same key, distinct events} (r1s);

    \draw[->,purple] (r1s) -- node[left]{ins(3,5)} (r1m);
    \draw[->] (r1m) -- (r1e);
    \draw[->,blue] (r2s) -- node[right]{ins(3,7)} (r2m);
    \draw[dashed,->,purple] (r1m) -- (r2s);
    \draw[dashed,->,blue] (r2m) -- (r1e);
  \end{scope}

  \begin{scope}[xshift=2.7cm]
    \coordinate[cpt,draw=white] (origin);
    \node[cpt,purple,label=left:{$\emptyset$}] (r1s) [below=2mm of origin] {};
    \node[cpt,purple,label=left:{$\{(3, 5)\}$}] (r1m) [below=5mm of r1s] {};
    \node[cpt,label=left:{$\{(3, 5),(3, 7)\}$}] (r1e) [below=5mm of r1m] {};

    \node[cpt,blue,label=right:{$\emptyset$}] (r2s) [right=\repgap of origin] {};
    \node[cpt,blue,label=right:{$\{(3, 7)\}$}] (r2m) [below=5mm of r2s] {};
    \node[cpt,label=right:{$\{(3, 5), (3, 7)\}$}] (r2e) [below=5mm of r2m] {};

    \path (origin) -- node [red,below=2.4cm] {same key, overlapping events} (r2s);

    \draw[purple,->] (r1s) -- node[left]{ins(3,5)} (r1m);
    \draw[->] (r1m) -- (r1e);
    \draw[blue,->] (r2s) -- node[right]{ins(3,7)} (r2m);
    \draw[->] (r2m) -- (r2e);
    \draw[dashed,->,purple] (r1m) -- (r2e);
    \draw[dashed,->,blue]  (r2m) -- (r1e);
  \end{scope}

  \begin{scope}[xshift=7.7cm]
    \coordinate[cpt,draw=white] (origina);
    \coordinate[cpt,draw=white,right=\repgap of origina] (originb);

    \node[cpt,purple,label=left:{$\emptyset$}] (r1s) [below=0mm of origina] {};
    \node[cpt,purple,label=left:{$\{(4, 5)\}$}] (r1m) [below=5mm of r1s] {};
    \node[cpt,label=left:{$\{(3, 7),(4, 5)\}$}] (r1e) [below=5mm of r1m] {};

    \node[cpt,blue,label=right:{$\emptyset$}] (r2s) [below=1mm of originb] {};
    \node[cpt,blue,label=right:{$\{(3, 7)\}$}] (r2m) [below=5mm of r2s] {};
    \node[cpt,label=right:{$\{(3, 7),(4, 5)\}$}] (r2e) [below=5mm of r2m] {};

    \path (origina) -- node [below=2.4cm] {different key, overlapping events} (originb);

    \draw[purple,->] (r1s) -- node[left]{ins(4,5)} (r1m);
    \draw[->] (r1m) -- (r1e);
    \draw[blue,->] (r2s) -- node[right]{ins(3,7)} (r2m);
    \draw[->] (r2m) -- (r2e);
    \draw[dashed,->,purple] (r1m) -- (r2e);
    \draw[dashed,->,blue] (r2m) -- (r1e);
  \end{scope}

\end{tikzpicture}
  \caption{Operations executing on the key-value store, across two replicas.}
  \label{fig:key-val-store}
\end{figure}
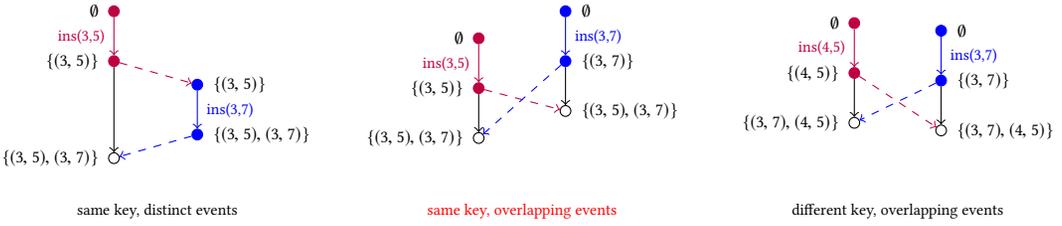

\paragraph{Example: key-value store} Consider a simple key-value store and an
{\em operation}, {\tt insert(k,v)}. When executed on a replica, the operation
tests (using a predicate {\tt present(k)} whether an entry with key {\tt k} is
already in the store. If so, the operation has no effect. Otherwise, it inserts
the pair {\tt (k,v)}. Furthermore, it issues an effect {\tt ins(k,v)} that
simply tells all the other replicas to execute the insertion as well, without
further tests. Without any other requirements on the store, this implementation
is eventually consistent and is an example of a CRDT. There are no conflicting
operations (as the {\tt insert} operation does not conflict with another
instance of itself). In Figure~\ref{fig:key-val-store}, there are three
executions, all eventually consistent.

Let us consider a store that has an invariant that all the entries must have
unique keys. Is this invariant maintained? The behavior of {\tt insert(k,v)}
depends on the value of {\tt present(k)}. But {\tt present(k)} is evaluated
based only on the information the replica has. Thus it is entirely possible that
another replica executes another {\tt insert(k,v)} operation with the same value
of {\tt k}, leading to a store with non-unique keys which violate the invariant.
Thus in this case the {\tt insert} operation can conflict with another instance
of itself, but only when both want to insert an entry with the same key.  In
Figure~\ref{fig:key-val-store}, the execution on the left and the one in the
center violate the invariant, while the execution on he right preserves it.

To ensure that the invariant is preserved, the developer writing the {\tt
insert(k,v)} operation introduces the consistency guard {\tt
presentR(k)==presentG(k)}, which requires that the replica value ({\tt
presentR(k)}) is equal to the global value {\tt presentG(k)}\footnote{The  language we introduce has a different syntax for specifying guards. For
brevity, here we provide directly the two state predicate that the guard
defines.}. The guard prevents other replicas from executing {\tt insert(k,v)}
with the same value of {\tt k} in parallel, as such executions would modify {\tt
presentG(k)} and thus invalidate the guard.  However, the guard does not prevent
parallel execution of {\tt insert(k',v)} for {\tt k'} different from {\tt k}.

\paragraph{Global state} The consistency guard refers
to a global state. This global state is never computed during the distributed
execution, but it is well-defined at each moment of the computation and the
guard (i.e., a relation between the global state and the replica state) can be
maintained.
The global state is defined using the {\em arbitration
order}~\cite{Burckhardt:Book} which is a total order on all events in a
computation. The arbitration order can be maintained in a standard way without
any synchronization. For a particular event in a computation, the global state
is obtained by evaluating all the effects that are before that event in the
arbitration order.

\paragraph{Replica state} During the computation, a
replica of course does not have access to the global state.  All it has is the
effects it has seen (note that there might be effects that the replica has not
seen yet that will be arbitrated before the current operation). Thus the replica
state is determined using the {\em visibility partial order} $\vis$: an effect
$e$ is after an effect $f$ in the visibility order iff the operation that
produced $e$ ran at a replica which has seen $f$ at that time. We require that
the arbitration order and the visibility order agree. This requirement is called
causal consistency and can be maintained without any blocking synchronization.

\paragraph{Maintaining the consistency guards} We are now ready to explain how
consistency guards are maintained. If a replica starts to execute an operation
guarded by a guard $g$ and producing an effect $\eta$, it makes sure that for
every other effect $\eta'$ either (i) $\eta$ and $\eta'$ were not produced in
parallel, i.e. $\vis(\eta,\eta')$ or $\vis(\eta',\eta)$, or (ii) $\eta'$ does
not invalidate $g$. That is, the operations that are allowed to run in parallel
do not invalidate $g$. Thus if $g$ is true when the operation starts, it is true
while the operation executes.  (We provide only an intuition here, see also
Section~\ref{sec:locks} for a stronger version of (ii) we need.)

This condition is possible to enforce by taking a distributed lock associated
with  $g$, and thereby disallowing  all conflicting operations (operations such
that their effects can modify the global state in a way that might invalidate
$g$) to run in parallel. Another replica considers the lock released when it
receives the effect of the operation that took the lock.

\paragraph{CARDs for the user} We show how our system satisfies the points (a)
to (f) above. Let us first consider the key-value CARD from the point of view of
the user.
\begin{compactitem}
  \item[(a)] Strong eventual consistency is achieved in a standard way by having
  the arbitration order. Each replica maintains a sequence of effects ordered by
  the arbitration order, so eventually the state at every replica will be
  obtained by evaluating the same effects in the same order.
  \item[(b)] Availability is achieved because operations are executed without
  blocking synchronization when possible. For instance, if {\tt k} is different
  from {\tt k'}, then {\tt insert(k,v)} and {\tt insert(k',v)} do not need to
  synchronize. Indeed, the effect {\tt ins(k',v)} does not invalidate the guard,
  as it does not change either the replica value or the global value of {\tt
  present(k)}.
 \item[(c)] Application invariants (despite the presence of conflicting operations) are maintained thanks to the consistency guards. We explained how the guard for {\tt insert} protects the invariant that the store contains entries with unique keys.
\end{compactitem}

\paragraph{CARDs for the developer} For the developer of a CARD, the following properties hold.
\begin{compactitem}
  \item[(d)] Modular reasoning: Consistency guards allow specifying the
  assumptions that a method relies on without considering what other methods
  might be operating on the same CARD. For instance, the guard for the {\tt
  insert(k,v)} ensures that the operation is correct regardless of what the
  other operations do.
  \item[(e)] Sequential reasoning: The consistency guards allow sequential
  reasoning about correctness of each individual operation, even though these
  operations run in a distributed system. The reason is that the guard is the
  only assumption that the operation makes on its distributed environment. We will provide an overview of the reasoning needed to prove correctness of an operation in Section~\ref{sec:overview}.
 \item[(f)] Algorithmic conflict detection: In our setting, the conflicts are
 between guards and effects. For instance, the guard {\tt
 presentR(k)==presentG(k)} is in conflict with the effect {\tt ins(k,v)}. Given
 consistency guards, we provide a weakest-precondition based algorithm that uses
 an SMT solver to automatically infer potential conflict between effects and
 guards at compile time.  We use the results to introduce necessary blocking
 coordination (with no unnecessary coordination). In particular, this means that
 such a system behaves as a CRDT in cases where the data structure supports
 conflicting operations, but they are never executed.
\end{compactitem}

\paragraph{Core calculus for CARDs} We introduce $\lambda^Q$, a core calculus
for specifying CARDs. It extends the $\lambda$ calculus by introducing terms for
queries (that create consistency guards) and for emitting effects. The calculus
generalizes the description above by allowing a replica to issue nested queries
(that impose one consistency guard each) before issuing an effect. The calculus
is typed using refinement (liquid) types that allow expressing pre- and
post-condition for each operation. Given an invariant $I$, we can prove it by
typechecking -- we can show that each operation typechecks with its pre- and
post- condition set to $I$.

\paragraph{Contributions.} To summarize, this paper makes the following contributions.
\begin{compactitem}
  \item We extend CRDTs to CARDs, allowing conflicting operations, and enabling
  programmers to modularly specify conflicts with consistency guards. [Section~\ref{sec:cards}]
  \item We introduce $\lambda^Q$, a core calculus
  for specifying CARDs. [Section~\ref{sec:lang}]
  \item We show that invariants on CARDs can be proved sequentially and
  modularly. To this end, we introduce a refinement (liquid) type system  for
  $\lambda^Q$ and show that it is a (sequential and modular) proof system for
  CARD invariants and more generally for correctness of CARD operations.
  [Section~\ref{sec:lang}]
  \item We provide a weakest pre-condition based algorithm for automatically
    inferring the minimal required synchronization between
    replicas in CARDs. [Section~\ref{sec:locks}]
  \item We describe a protocol that implements CARDs and prove it correct.
  [Section~\ref{sec:replica}]
  \item We implement the automated conflict inference algorithm and evaluate it
  on several small, but representative replicated data-types. The results show
  that the inference needed to detect conflicts between operations is well
  within the scope of current SMT solvers. [Section~\ref{sec:evaluation}]
\end{compactitem}


\section{Writing and Verifying CARD Operations}
\label{sec:overview}

We provide an overview of \CARDs and $\lambda^Q$ on an illustrative
example: a bank account where some operations conflict with each
other.
We explain how the application is programmed with $\lambda^Q$
operations over a general-purpose CARD, and show how static conflict
information can be inferred for the CARD and used to verify
application-specific properties for the bank account.
We then extend the example to show how non-commuting effects can be handled.
The application consists of {\tt withdraw} and {\tt deposit} operations over a
\texttt{Counter} CARD (simple integer value that supports addition and
subtraction). Executing these operations at a replica emits Counter effects
which will eventually by processed by other replicas.

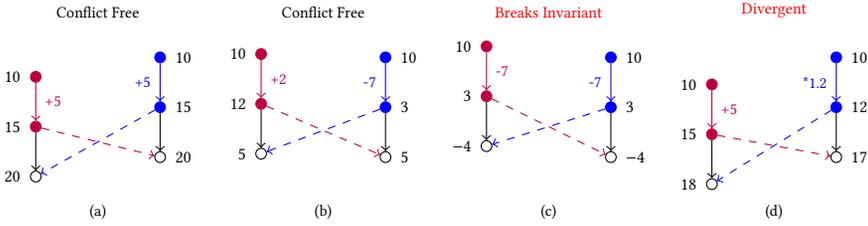
\begin{figure}
  \begin{tikzpicture}[font=\tiny]

  \tikzset{cpt/.style={draw=black,fill=white,circle,inner sep=.5mm}}
  \def\repgap{1.5cm}

  \foreach \id\si\oa\ob\ma\mb\resa\resb\label\shifta [count=\i] in
  {
    a/10/+5/+5/15/15/20/20/Conflict Free/1mm,
    b/10/+2/-7/12/3/5/5/Conflict Free/-2mm,
    c/10/-7/-7/3/3/-4/-4/\color{red}Breaks Invariant/-3mm,
    d/10/+5/*1.2/15/12/18/17/\color{red}Divergent/2mm
  } {
    \begin{scope}[xshift=(\i*3cm)]
      \coordinate[cpt,draw=white] (origin);
      \node[cpt,purple,label=left:{$\si$}] (r1s) [below=\shifta of origin] {};
      \node[cpt,purple,label=left:{$\ma$}] (r1m) [below=5mm of r1s] {};
      \node[cpt,label=left:{$\resa$}] (r1e) [below=5mm of r1m] {};

      \node[cpt,blue,label=right:{$\si$}] (r2s) [right=\repgap of origin] {};
      \node[cpt,blue,label=right:{$\mb$}] (r2m) [below=5mm of r2s] {};
      \node[cpt,label=right:{$\resb$}] (r2e) [below=5mm of r2m] {};

      \path (origin) -- node [above=4mm] {\label} (r2s);
      \path (origin) -- node [below=1.8cm] {(\id)} (r2s);

      \draw[purple,->] (r1s) -- node[right]{\oa} (r1m);
      \draw[->] (r1m) -- (r1e);
      \draw[blue,->] (r2s) -- node[left]{\ob} (r2m);
      \draw[->] (r2m) -- (r2e);
      \draw[purple,dashed,->] (r1m) -- (r2e);
      \draw[blue,dashed,->] (r2m) -- (r1e);
    \end{scope}
  }
\end{tikzpicture}

\caption{Examples of conflict-free and conflicting operations in executions
in a bank account with three operations: deposit ($+$), withdraw ($-$), and
interest ($*$). Each execution consists of two replicas, each executing one
operation with no coordination (solid line), and then broadcasting their
effects (dashed line). Executions (a) and (b) are conflict-free, (c) produces
a negative balance (breaking an application invariant), and (d) leads to a
divergent state (breaking SEC).}
  \label{fig:noncom-effects}
\end{figure}

\paragraph{Problem and desired result}
The bank account has three requirements: {\em strong eventual
  consistency}, {\em availability}, and preserving {\em
  application-specific invariant} $I$: the bank account value should
never be negative.
The \texttt{Counter} effects produced by {\tt deposit} and {\tt
  withdraw} ($\mathtt{Add}\;n$ and $\mathtt{Sub}\;n$, respectively)
commute, and thus SEC can be achieved without damaging availability
(as in CRDTs).
However, the replicas need to coordinate in order to maintain the
invariant $I$.
The \texttt{withdraw} operation can be made ``smart'' so that it
decides not emit a $\mathtt{Sub}\;n$ effect if it sees that the
account is too small, but if for example two $\mathtt{withdraw}\; 7$
operations running on separate replicas see a store value of $10$ and
make their decisions before they see each other, they will together
reduce the account to $-4$, breaking the invariant anyway
(See Figure~\ref{fig:noncom-effects}c).
Thus two {\tt withdraw}s cannot run in parallel; if they do, their
safety logic might not work.
On the other hand, multiple {\tt deposit}s can run in parallel, and
even multiple {\tt deposit}s and a single {\tt withdraw} can run in
parallel.
The desired technique should therefore statically detect a conflict between the
two {\tt withdraw}s, and (i) avoid this conflict, while (ii) allowing all other
operations run in parallel without incurring a performance penalty (and thus
preserve availability to the extent possible).

A CARD $D$ is a rich datatype consisting of a basic store type $S(D)$,
a type $E(D)$ of \emph{effects} which transform the store type, and a
type $C(D)$ of \emph{consistency guards} that state conditions of
partial equivalence between store values.
For example, a consistency guard on a list CARD might state that two
list values are identical up to some nth element.
We use guards in CARD applications to state what kind of consistency
is required (and thus what kind of interference is disallowed) for a
particular access of replicated store data.

\begin{figure}[h]
  {\footnotesize
  \begin{align*}
    S(\mathtt{Counter})&\defn\; \mathbb{Z} & \denote{\mathtt{Add}\;n}
    &\defn\; \lambda x. \; x + n & \denote{\top} &\defn\; \top \\
    E(\mathtt{Counter})&\defn\; \mathtt{Add} \; \mathbb{N} \;|\;
                         \mathtt{Sub} \; \mathbb{N} \;|\;
                         \mathtt{Set} \; \mathbb{Z} & \denote{\mathtt{Sub}\;n}
    &\defn\; \lambda x. \; x - n & \denote{\mathtt{LE}} &\defn\; s_r \le s_g \\
    C(\mathtt{Counter})&\defn\; \top \;|\;
                         \mathtt{LE} \;|\;
                         \mathtt{GE} \;|\;
                         \mathtt{EQ} &
                                       \denote{\mathtt{Set}\;n}
    &\defn\; \lambda x. \; n & \denote{\mathtt{GE}} &\defn\; s_r \ge
                                                      s_g \\
    & & & & \denote{\mathtt{EQ}} &\defn\; s_r = s_g \\
  \end{align*}
  \caption{Definition of the \texttt{Counter} CARD}
  \label{fig:def-counter}
  }
\end{figure}

The example CARD we are using here is the \texttt{Counter}, defined in
Figure~\ref{fig:def-counter}, which uses an integer as its store type,
supports simple numerical effects, and provides lower ($\mathtt{LE}$)
and upper ($\mathtt{GE}$) bound guard measures.
Having defined this datatype, we can automatically
infer the complete set of conflict relationships between the effects
and guards up front without needing to know what varying
application-specific safety properties they will be used to implement.

\paragraph{Operations}
We define \emph{operations} over a CARD $D$ using $\lambda^{Q}$, an
extension of the $\lambda$-calculus.
An operation is a program which runs an effect and/or returns
information to the caller based on partial knowledge of the store's
current value.
For example, consider the \texttt{withdraw} operation for our bank
account example written in $\lambda^Q$:
\[
  \mathtt{withdraw}\defn \lambda n. \;
  (Q \; \mathtt{LE} \triangleright x. \;
  (\mathtt{if} \; (x \ge n) \;
  \mathtt{then} \; R.(\mathtt{Sub}\;n,n) \;
  \mathtt{else} \; R.(\mathtt{NoOp},0)))
\]
The term $Q \; \mathtt{LE} \triangleright x. \; (\ldots)$
binds a snapshot of the store to $x$ for use in the if-expression.
In order to choose safely whether to subtract the argument value $n$
from the store, the snapshot bound to $x$ must not be greater than the
current store value.
Thus we annotate the term with the $\mathtt{LE}$ guard to declare
that the current store must be less than or equal to the value we bind
to $x$ -- this safely under-approximates the condition that $n$ should
be at most the current store value.
The base term $R.\;(e,a)$ adds $e$ as an effect to the store and
returns $a$ to the caller.
In our case, we only add the $\mathtt{Sub}\;n$ to the store if we know
that it is safe, and we return the value we decided to subtract (if
any) to the caller.
A reader familiar with the challenges of distributed systems might be
suspicious of this ``current value'' for the replicated store.
We will define precisely what this means in Section~\ref{sec:lang}.

Notice that in writing this safe operation, we did not explicitly
declare conflicts with other operations or said anything about event
orderings.
A replica running \texttt{withdraw} uses the conflict information
previously generated for \texttt{Counter} to impose the network
ordering constraints needed to enforce our \texttt{LE} guard.

\subsection*{Checking a Dependent Operation Type}

Because guards reduce the concurrent problem of operation correctness
to a sequential one, we can use standard sequential reasoning tools to
verify operation behavior.
In particular, we extend the type inference rules of Liquid
Types~\cite{liquidtypes} to cover $\lambda^Q$'s unique terms.
Operations are then type-checked with respect to a specification on
the behavior of the event they produce.
For example, the specification we check for the \texttt{withdraw}
operation states formally the behavior we described earlier:
\[
  \mathtt{withdraw} \; : \; (n:\mathtt{Nat}) \to
  \mathtt{Op}(\mathtt{Counter},\mathtt{Int},(s \ge 0 \Rightarrow s'
  \ge 0) \land (a = s - s'))
\]
This operation type states that \texttt{withdraw}, given a natural
number amount, is an operation over \texttt{Counter} returning an
integer and meeting two refinement conditions:
\begin{inparaenum}
\item the bank account's non-negative invariant is preserved and
\item the return value ($a$) reflects exactly the amount that is
  removed from the account.
\end{inparaenum}
The $\mathbf{a}$, $\mathbf{s}$, and $\mathbf{s'}$ in the specification
are special free variables used to refer to the return value, the
store value before applying the operation's effect, and the store
value after the operation completes.
Our typing rules will reduce this to a Liquid Type which must be
checked.
The argument $n$'s type \texttt{Nat} is itself an example of a Liquid
Type which we will use in the derivation.
\[
  \denote{n: \mathtt{Nat}} = \denote{n:\{\nu : \mathtt{Int} \;|\; \nu
    \ge 0\}} = n \ge 0
\]

We now check the operation type against our \texttt{withdraw}
definition.
The correctness of \texttt{withdraw} depends on the store value
guarantee it demands via the \texttt{LE} query guard, and the
\textsc{type_q} typing rule adds that guarantee to the context.

\ottusedrule{\ottdruletypeXXq{}}

Thus typing the outer term
$Q \; \mathtt{LE} \triangleright x. \; \mathtt{if}\ldots$ adds
$x:\{\nu : \mathtt{Int} \;|\; \nu \le s\}$ which states that the value
bound to $x$ is less than or equal to the pre-effect store value.
\[
  \denote{x:\{\nu : \mathtt{Int} \;|\; \nu \le s\}} = x \le s
\]
Following the positive branch of the
$\mathtt{if}\;(x \ge n)\;\mathtt{then}\{\ldots\}\mathtt{else}\{\ldots\}$
further adds $x \ge n$ to the context.
We arrive at the final constraint-solving problem by applying the rule

\ottusedrule{\ottdruletypeXXr{}}

\noindent to the $R.\; (\mathtt{Sub}\;n,n)$ base term that gives the effect and
return value that a successful \texttt{withdraw} produces.
Following the \textsc{type_r} rule, we need to show
\[
  \Gamma \vdash n : \{\nu : \text{Int} \;|\; s'=((\lambda s.\; s -
  n)(s)) \Rightarrow (s \ge 0 \Rightarrow s' \ge 0) \land (a = s -
  s')\}
\]
to finish checking the positive \texttt{then} branch, which becomes
the simple constraint problem
\[
  (n \ge 0) \land (x \le s) \land (x \ge n) \land (s' = s - n)
  \Rightarrow (s \ge 0
  \Rightarrow s' \ge 0) \land (a = s -
  s')
\]
when $\denote{\Gamma}$ is unpacked according to the Liquid Type rules.
The trivial \texttt{else} branch check is clearly satisfied by the
fact that its effect does nothing.
\[
  \denote{\Gamma} \land (s' = s) \Rightarrow (s \ge 0 \Rightarrow s'
  \ge 0) \land (a = s - s')
\]

\subsection*{CARDs with Non-Commutable Effects}

Many replicated data reasoning models and implementations require all
effects on the replicated store to be commutable in order to simplify
the way histories are merged.
In the interest of generality, CARDs do allow non-commuting store effects, and
our reasoning technique and implementation technique are equipped  to handle
them efficiently.
To demonstrate this flexibility and build some more intuition, let's
take a look at some example applications. More examples can be found
in Section~\ref{sec:evaluation}.
Figure~\ref{fig:noncom-effects}d illustrates how non-commutative effects
(here, $+5$ and $\times1.2$) can lead to replicas diverging,
violating strong eventual consistency.

\subsubsection*{Bank Account with Interest and Non-commuting Effects}
\label{sec:bank-with-interest}

An obvious challenge of non-commutable effects is maintaining SEC.
Our approach, following~\cite{Burckhardt:2012}, is to use an {\em
  arbitration order}, which is a total order on events which a replica
chooses to evaluate the current value.
The key is that the arbitration order must be chosen and maintained
consistently across replicas.
Such an order can be maintained using a standard combination of Lamport
clocks and replica identifiers and by inserting newly received updates
appropriately in history instead of appending them.

We now extend our example to show that even with non-commuting
effects, strong eventual consistency can be achieved without blocking.
Consider our bank account over an extended CARD \texttt{Counter'} with
new effect $\denote{\mathtt{Interest}}\defn \lambda s. s * 1.2$, and
suppose we write a new operation \texttt{safeBalance} which returns a
value that is definitely not less than the account's actual value.
\[
  \mathtt{safeBalance} : \mathtt{Op}(\mathtt{Counter'},\mathbb{Z},s' =
  s \land a \le s)
\]

The order of the {\tt Sub} and {\tt Interest} events matter, i.e., the
effects do not commute.
Most approaches~\cite{CRDTs,RedBlue} would declare these two
operations in conflict, and thus would be either disallowed (CRDTs) or
declared strongly consistent (RedBlue).
Furthermore, if effects are reordered at replicas, maintaining
guarantees about the relationship between the return value and the
global state becomes hard --- so using an operation that reads this
shifting state might require coordination.

However, the guard of {\tt safeBalance} allows us to infer that its
requirement does not conflict with either {\tt deposit} or {\tt
  interest}, so all three operations can be executed in parallel.
Because the desired behavior of \texttt{safeBalance} was verified
entirely based on its query guard, we can be sure that its behavior
survives effect reorderings.
Thus we achieve efficiency, even while ensuring application
properties, by depending on the arbitration order rather than
coordination to maintain SEC even with non-commutable effects.

\subsubsection*{Joint Bank Account and Chained Conflicts}

We have explained how using the arbitration order allows achieving SEC.
The downside is that due to non-commuting effects, detecting conflicts is in
general more difficult than it was for our first bank account example.  There
may exist effects which cannot violate a guard, but instead can change the
behavior of a non-commuting effect that does have the ability to violate a
guard.

To demonstrate, we extend the example to a bank account which is
jointly owned by two users, in which a user must first request a
withdraw (via {\tt request}) and wait for someone else to approve (via
\texttt{approve}) before actually performing it.

We use a $(\mathtt{Counter},\mathtt{Bool},\mathtt{Bool})$ tuple as the
store, which supports the effects and guards of the $\mathtt{Counter}$
as well as effects and guard
\begin{align*}
  \vspace{-4ex}
  \denote{\mathtt{Request}}&\defn \lambda (s,b_1,b_2).(s,\top,b_2) &   \denote{\mathtt{App?}}&\defn b_2(s_g) = b_2(s_r) \\
  \denote{\mathtt{Approve}}&\defn \lambda (s,b_1,b_2).(s,b_1,b_1) \\
  \denote{\mathtt{Reset}}&\defn \lambda (s,b_1,b_2).(s,\bot,\bot)
  \vspace{-4ex}
\end{align*}
in which \texttt{App?} guarantees that the second boolean seen has the
same value as the second boolean on the global store.

In this case, a user must first request a withdraw (via {\tt Request})
and wait for someone else to approve (via \texttt{Approve}) in order
for the withdrawal to have an effect.
\begin{align*}
  \mathtt{withdrawJ}\defn Q\;\mathtt{LE}\land\mathtt{App?}
                      \triangleright (s,b_1,b_2).(&\mathtt{if}\;(s \ge n
                      \land b_2)\;\\ &\mathtt{then}\;R.(\mathtt{Sub}\;n
                      \circ \mathtt{Reset},n)\;\\ &\mathtt{else}\; R.(\mathtt{NoOp},0))
\end{align*}
The operation {\tt withdrawJ} is guarded by $\mathtt{App?}$ to be sure
that the actual withdrawal of funds happens only if it was approved.
The operation {\tt withdrawJ} must not be concurrent with itself (as
before), but it is now also in conflict with anything that emits {\tt
  Approve}, as {\tt Approve} can invalidate $\mathtt{App?}$.

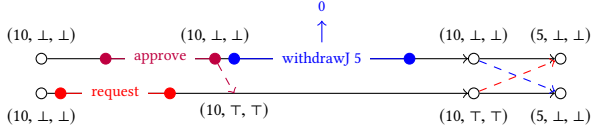
\begin{figure}
  { 

\def\repgap{3mm}
\def\oplen{2.8cm}
\def\synclen{1cm}

\begin{tikzpicture}[font=\tiny]
  \tikzset{cpt/.style={draw=black,fill=white,circle,inner sep=.5mm}}

  \node[cpt,label={$(10, \bot, \bot)$}] (r1s) {};
  \node[cpt,label={$(10, \bot, \bot)$}] (r1o) [right=(2*\oplen) of r1s]{};
  \node[cpt,label={$(5, \bot, \bot)$}] (r1e) [right=1cm of r1o]{};

  \node[cpt,label=below:{$(10, \bot, \bot)$}] (r2s) [below=\repgap of r1s] {};
  \node[cpt,label=below:{$(10, \top, \top)$}] (r2o) [below=\repgap of r1o] {};
  \node[cpt,label=below:{$(5, \bot, \bot)$}] (r2e) [below=\repgap of r1e] {};

  \draw[->] (r1s) -- (r1o);
  \draw[->] (r1o) -- (r1e);

  \draw[->] (r2s) -- (r2o);
  \draw[->] (r2o) -- (r2e);
  \draw[red,dashed,->] (r2o) -- (r1e);

  \node[cpt,purple] (r1o1a) [right=7mm of r1s] {};
  \node[cpt,purple,label=above:{$(10, \bot, \bot)$}] (r1o1b) [right=13mm of r1o1a] {};
  \draw[purple] (r1o1a) -- node[fill=white] {approve} (r1o1b);

  \node[cpt,blue] (r1o2a) [right=1mm of r1o1b] {};
  \node[cpt,blue] (r1o2b) [left=7mm of r1o] {};
  \draw[blue] (r1o2a) -- node[fill=white] (r1o2) {withdrawJ 5} (r1o2b);
  \draw[blue,->] (r1o2) -- node[at end,above] {0} ++(0, 5mm);
  \draw[blue,dashed,->] (r1o) -- (r2e);

  \node[cpt,red] (r2oa) [right=1mm of r2s] {};
  \node[cpt,red] (r2ob) [right=13mm of r2oa] {};
  \draw[red] (r2oa) -- node[fill=white] {request} (r2ob);

  \draw[purple,dashed,->] (r1o1b) -- node[at end,below,black]{$(10, \top, \top)$} (r1o2a|-r2oa);
\end{tikzpicture}

}
  \caption{Chained conflicts causing a problem. {\tt withdrawJ} saw an {\tt approve}, but the {\em approve} did not have an effect, since {\tt approve} did not see a {\em request}. So {\tt withdrawJ} failed and reported $0$ to the client. However, later a {\tt request} was arbitrated before {\tt approve} changing it effects, and making the execution of {\tt withdrawJ} invalid.}
  \label{fig:chained-conflicts}
\end{figure}

Now note that {\tt Approve} and {\tt Request} are non-commuting: the
behavior of {\tt Approve} is changed by a {\tt Request} existing
before it.
Consider a situation (illustrated in Fig. ~\ref{fig:chained-conflicts}) where
replica $r_1$ emits {\tt Approve} and then runs {\tt withdrawJ}, while
concurrently, replica $r_2$ emits {\tt Request}.
Let us assume that the arbitration order will eventually put the {\tt
  Request} before the effect of {\tt Approve}.
Then an execution can look as follows: replica $r_1$ sees an {\tt Approve}
(which does not set $\app$ to {\tt true} as there is no request pending) and
then $r_1$ executes a {\tt withdraw} while guaranteeing that there are no
concurrent $\mathtt{Sub}\;n$ or {\tt Approve} effects.
However, when the {\tt Request} from replica $r_2$ is received by
$r_1$, and the arbitration causes this effect to be ordered before the
\texttt{Approve}, then suddenly the behavior of the \texttt{Approve}
changes: it sets the second boolean to true.

Note that at the time of execution of {\tt withdraw}, the guard
\texttt{App?} would hold; however, the arrival of the {\tt Request}
and consequent re-evaluation of \texttt{Approve} would retroactively
invalidate the guard.
Thus \texttt{App?} must be in conflict with not just {\tt Approve},
but also with {\tt Request}, as it changes the behavior of {\tt
  Approve}, potentially causing violation.  We provide an algorithm
that finds such chained conflicts in Section~\ref{sec:locks-minimal}.

\sectionnewpage
\section{Conflict-Aware Replicated Datatypes}
\label{sec:cards}

We define CARDs, an abstract model of replicated data stores, and
executions based upon them.

\subsection{CARDs}

A \emph{conflict-aware replicated datatype} is a tuple $D = (S,E,C)$
where $S$ is the store type, $E$  is the type of \emph{effects}, and $C$
is the type of \emph{consistency guards}.
Effects and consistency guards are detailed below.
Informally, effects are store transformers and consistency guards
specify the exact semantic restrictions on consistency under which each
operation may execute under.
The key point behind CARDs is to automate the reasoning about the
interaction between effects and consistency guards.
This allows a developer to program CARD operations modularly, letting
the system handle conflicts in an automated manner.

\paragraph{CARD Effects}
The type $E$ is the type of \emph{effects} on the store.
A value $e:E$ has a denotation $\denote{e}$ which is an $S \to S$
function modifying a store value.

\begin{example}
  \label{ex:effects}
  In the bank account example, we have the effect type $E \defn
  \mathtt{Add}\;\mathtt{Nat} \;|\; \mathtt{Sub}\;\mathtt{Nat}$.
  Each effect is of the form $\mathtt{Add}\;n$ or $\mathtt{Sub}\;n$ for
  some positive integer $n$.
  The denotations of $\mathtt{Add}\;n$ and $\mathtt{Sub}\;n$ are given
  by $\lambda s .\; s + n$ and $\lambda s .\; s - n$, respectively.
\end{example}

\paragraph{Consistency Guards}
The type $C$ is the type of \emph{consistency guards} on the store
which describe measures of ``accuracy'' for partial knowledge of the
store value.
Consistency guards are \emph{semantic in nature}, i.e., they do not restrict the
ordering of operations like traditional consistency models (e.g., sequential
consistency, etc), but instead semantically restrict the updates to the store.
Formally, a value $c:C$ has a denotation $\denote{c}$ which is a two-state
predicate (of type $S \times S \to \mathbb{B}$) relating the ``global store
value'' ($s_g$) and a ``local store view'' ($s_r$) that some replica has.
We will write $c(s_1,s_2)$ to mean $[s_1/s_g][s_2/s_r]\denote{c}$.
We restrict all guards to be reflexive, as in $\forall s. c(s,s)=\top$ -- a
replica store view equal to the global store value represents complete
knowledge of the store.
Replicas and local store views are described fully in
Section~\ref{sec:replica}.

\begin{example}
  \label{ex:guards}
  In the running bank account example, the denotation of consistency guards have
  type $\mathtt{Int} \times \mathtt{Int} \to \mathbb{B}$.
  The guard $\mathtt{LE} \defn s_g \geq s_r$ restricts the global store
  value to be at least as great as the local store value.
  Intuitively, we will use the guard $\mathtt{LE}$ to ``guard'' withdraw
  operations --  any replica executing a withdraw operation will have a
  local store value that is at most the global value, ensuring that the
  withdraw does not decrease the balance below $0$.
  Informally, this implies that we need to restrict the global value
  from being decreased by other withdraw operations once the local
  replica has decided on a value of balance for the current withdraw
  operation.
  Another guards we will use in the bank account examples is $\mathtt{EQ}  \defn s_g = s_r$.
\end{example}

\paragraph{Effect Classes}
A CARD's effect type $E$ will often generate an infinite set of effect
values.
For example, the Counter CARD includes an $\mathtt{Add}\;n \defn
\lambda s .\; s + n$ effect for all $n:\mathbb{N}$.
In order to facilitate automated reasoning about effects and guards
that is necessary for runtime locking decisions, we assume that this set
of infinite effects are divided into a finite set $\overline{E}$ of
\emph{parametric effect classes}.
The choice of classes must be made by the developer of the CARD, and
is most effective when each class is characterized by the relationship
to the set of relevant guards.
In our examples, we assume that the type $E$ is a non-recursive algebraic data
type, with values of each type variant being one class.
We will elide this classification detail for the rest of the paper;
when an algorithm quantifies $\forall e:E$ we assume that we are using
a finite $E$ or a quantification over the finitely many parametric classes of
$\overline{E}$.

\begin{example}
  \label{ex:effect-classes}
  For the bank account example, the obvious choice is to classify
  effects by constructor: $\overline{E} := \{ \mathtt{Add^\forall},
  \mathtt{Sub^\forall} \}$ where $\mathtt{Add^\forall}$ and
  $\mathtt{Sub^\forall}$ include events of the form $\mathtt{Add}\;n$
  and $\mathtt{Sub}\;n$, respectively.
  Each effect in the effect class behaves similarly with respect to the
  guards $\mathtt{LE}$ and $\mathtt{EQ}$.
  For example, all $\mathtt{Sub}\;n$ effects may cause the condition
  $\mathtt{LE} \defn s_g \geq s_r$ to be violated if the global store is
  updated with it, while $\mathtt{Add}\;n$ cannot cause the same.
\end{example}

\subsection{CARD Executions}
\label{sec:cards-executions}

Following standard practice
(see~\cite{Burckhardt:Book,Burckhardt:2012,Burckhardt:2014,Gotsman:2016}),
we describe the execution history of an eventually consistent
replicated store using a set of \emph{events} that each represent the
execution of a single operation on the data store.
Events contain an effect that changes the store and a return value
that gives some information about the store back to the caller.
In addition, CARD events contain a set of \emph{active guards} that represent
the semantic consistency restrictions on the event.
Events are ordered by an \emph{arbitration total order} in order to
support CARDs with non-commutable effects.
Such an order must be decided consistently by all members of the
replicated store without coordination -- time stamps and Lamport
clocks can be used for this purpose, or it can be omitted in
implementation for systems which only make commutable store updates.

\paragraph{Active Guards}

Each event has a set of (zero or more) \emph{active guards}, (or AGs for short).
An event's AGs represent consistency guards that a replica had when producing
the effect.
Since we will allow a replica to impose a series of consistency guards to
produce one effect, each event might have more than one AG.
Associated with each AG is the subset of previous events that the
replica witnessed when it imposed the consistency guard.
This encodes the standard visibility relation between an AG and an event.
The AG is also associated with the consistency guard it represents.

\paragraph{$D$-executions}

Formally, a \emph{$D$-execution} for a CARD $D$ is a tuple
$L=(s_0,W,G,\grd,\ar,\vis)$ where:
\begin{itemize}
\item $s_0:S(D)$ is the initial store value
\item $W$ is a finite set of events.
\item $G$ is a finite set of active guards.
\item $\grd : W \to \mathbb{P}(G)$ gives the set of AGs for an event.
  Every AG is associated with a single event, which we denote by
  $\grd^{-1} : G \to W$.
\item $\ar \subseteq (W \times W)$ is the arbitrary total ordering on
  events.
\item $\vis \subseteq (W \times G)$ is our guard-based visibility
  relation, which indicates whether an AG witnesses an event.  We
  denote by $\vis^{-1} : G \to \mathbb{P}(W)$ the set of all events
  witnessed by an AG.
\end{itemize}
A $D$-execution also defines the following functions for examining
events and active guards:
\begin{itemize}
\item $\eff: W \to E(D)$ gives the $D$-effect an event holds
\item $\rval: W \to A$ gives the return value (of some type $A$) an
  event holds
\item $\gc : G \to C(D)$ gives the consistency guard an active guard
  was formed from.
\end{itemize}

\begin{example}
  \label{ex:events}
  In our running bank account example, two instances of events can be:
  \begin{compactitem}
  \item A withdraw event $\eta_w$ with effect
    $\eff(\eta_w)=\lambda s.\; s - 10$ reducing the store by $10$
    while returning the value $\rval(\eta_w)=10$ and guarded by a
    singleton active guard set $\grd(\eta_w)=\{g\}$ which maintains
    consistency guard $\gc(g)=\mathtt{LE}$ for the store with respect
    to $100$, the store value it witnessed when $\eta_w$ was being
    created.
  \item A deposit event $\eta_d$ with effect
    $\eff(\eta_d)=\lambda s.\; s + 100$ indicating that the effect of
    the event increases the store value by $100$, while returning
    $\rval(\eta_d)=100$, and being (not) guarded by an empty set
    $\grd(\eta_d)=\emptyset$ indicating that the replica made no store
    queries when creating $\eta_d$.
  \end{compactitem}
\end{example}

\paragraph{Evaluations}
The \emph{store evaluation} of a $D$-execution $L$, written as
$\eval(L)$ is the store value arrived at by starting with $s_0$ and
applying $\eff(\eta_i)$ for each $\eta_i \in W$ in $\ar$ order.
Formally, if $W = \{ \eta_0, \eta_1, \ldots \eta_n \}$ with each
$i < j \implies \ar(\eta_i, \eta_j)$, then
$\eval(L) = (\denote{\eff(\eta_n)} \circ \denote{\eff(\eta_{n-1})}
\cdots \denote{\eff(\eta_0)})(s_0)$.

\begin{example}
  \label{ex:execs}
  Continuing Example~\ref{ex:events}, given a $D$-execution
  $L = (0, \{ \eta_w, \eta_d \},G,\grd,\ar,\vis)$ where
  $\ar(\eta_d, \eta_w)$, the store evaluation $\eval(L)$ is given by
  $((\lambda s.\; s - 10 \;\circ\; \lambda s.\; s + 100)) (0)$, i.e.,
  $90$.
\end{example}

\begin{definition}[sub-executions]
  We define a \emph{sub-execution} of a $D$-execution
  $L=(s_0,W,G,\grd,\ar,\vis)$ as any other $D$-execution
  $L'=(s_0,W',G',\grd',\ar',\vis')$ for which $W' \subseteq W$,
  $G' \subseteq G$, $\grd' \subseteq \grd$, $\ar' \subseteq \ar$,
  $\vis' \subseteq \vis$, and
  $\forall \eta \in W'.\; \grd(\eta)=\grd'(\eta)$ (so that any
  remaining event retains all it's active guards).
\end{definition}

The above definition says that for $L'$ to be a sub-execution, $W'$ must retain
any event that is visible to any guards remaining in $G'$ (and thus which has
``caused'' any observable effect).

\paragraph{Pre-Executions}
We define the \emph{pre-execution} of an event $\eta$ in a
$D$-execution $L$ as the sub-execution of $L$'s components to the events
ordered by $\ar$ before $\eta$, and we
write this as $L_{\eta}$ for short.
The \emph{pre-store} of $\eta$ is then the evaluation of $L_{\eta}$,
and the \emph{post-store} is $\denote{\eff(\eta)}(L_{\eta})$.
In further discussion, the \emph{global store value} when an operation
is being executed at a replica, refers to the pre-store value in the
abstract execution (as per the arbitration order).
Note that this global store value is not stored explicitly, and the
replica executing an operation cannot learn the global store value
without additional coordination with other replicas.

\begin{example}
  \label{ex:pre-execs}
  Continuing Example~\ref{ex:execs}, the pre-execution of $\eta_w$ is
  given by
  $L_{\eta_w} = (0, \{ \eta_d
  \},\emptyset,\emptyset,\emptyset,\emptyset)$.  The pre-store and
  post-store values are $100$ and $90$, respectively.
\end{example}

Similarly, we define the \emph{vis-execution} of a guard $g$ in a
$D$-execution $L$ as the pre-execution of $L$'s components to the events
in $\vis^{-1}(g)$, and we write this as
$L_g$ for short.
The \emph{vis-store} of $g$ is then the evaluation of $L_g$.

\paragraph{Well-Formed Executions}
We consider a $D$-execution \emph{well-formed} if all of the following
hold:
\begin{enumerate}
\item An event's AGs can only be influenced by other events which are
  preceding ($\ar$ respects $\vis$, causal consistency), i.e.,
  $
    \forall \eta_1 \in W.\; \forall g \in \grd(\eta_1).\; \forall
    \eta_2 \in \vis^{-1}(g).\; \ar(\eta_1,\eta_2)
  $
\item All AGs are satisfied, meaning that their pre-store and
  vis-store satisfy their consistency guard (guard-compliance), i.e,
  $
    \forall \eta \in W.\; \forall g \in \grd(\eta).\;
    \gc(g)(\eval(L_{\eta}),\eval(L_g))
  $
\item An AG that sees an event also sees the preceding events seen by
  that event's AGs (transitivity of $\vis$), i.e.,
  $
    \forall \eta_1,\eta_2 \in W.\; \forall g_2,g_3 \in G.\;
    \vis(\eta_1,g_2) \land g_2 \in \grd(\eta_2) \land \vis(\eta_2,g_3)
    \Rightarrow \vis(\eta_1,g_3)
  $
\end{enumerate}

\paragraph{Event Specifications}
We specify correctness of events using constraints on the
relation between the pre-store value $s$ before the execution of the
event, the post-store value $s'$ after the execution of the event, and
the return value $a$ associated with the event.
Formally, an \emph{event specification}  is a predicate $\varphi$ of
type $S \times S \times A \to \mathbb{B}$.

\begin{definition}[Satisfaction of an Event Specification]
  \label{def:event-sat}
  An event $\eta$ in an execution $L$ \emph{satisfies} a specification
  $\varphi$, written $\eta \models_L \varphi$, iff $\varphi$ holds
  for $\eta$'s pre-store as $s$, $\eta$'s post-store as $s'$ and
  $\eta$'s return value as $a$.
  \[
    \eta \models_L \varphi \Leftrightarrow s=\eval(L_{\eta}) \land
    s'=\denote{\eff(\eta)}(s) \land a=\rval(\eta) \Rightarrow \varphi(s,s',a)
  \]
\end{definition}

\begin{example}
  \label{ex:event-specs}
  For the running bank account example, we may want the properties that
  \begin{inparaenum}[(a)]
  \item the post-store value is non-negative, and
  \item the change in the store value is equal to the return value of
    each event.
  \end{inparaenum}
  The event specification $\varphi(s, s', a) \defn s' \geq 0 \land s -
  s' = a$ exactly states this specification.
  Both the events $\eta_w$ and $\eta_d$ satisfy this specification: for
  example, in the case of $\eta_w$, we have $\psi \land s' = e(s) \defn s \geq
  100 \land s' = s - 10 \implies s' \geq 0 \land s - s' = 10 \defn \varphi$.
\end{example}

In Section~\ref{sec:lang}, we describe $\lambda^Q$, a programming
language for writing CARD \emph{operations}, programs that dynamically
produce an event based on a replicas (limited) knowledge of the current
store value.
The operational semantics of $\lambda^Q$ operations only produce well-formed
executions (Theorem~\ref{thm:op-rules-wf}). The type system of $\lambda^Q$ can
be used to check that an operation only produces events which satisfy a
particular specification (Theorem~\ref{thm:prod-event-sat}). This property makes
proving invariants straightforward (Theorem~\ref{lem:invariants}).


\sectionnewpage
\section{Language and Type System for CARD Operations}
\label{sec:lang}

In this section we describe the syntax, operational semantics, and
refinement typing rules for $\lambda^Q$, a core calculus language
extending the CBV $\lambda$-calculus for defining CARD operations.

\subsection{CARD Operations}

\begin{figure}
  \centering
  \includegraphics[width=0.8\textwidth]{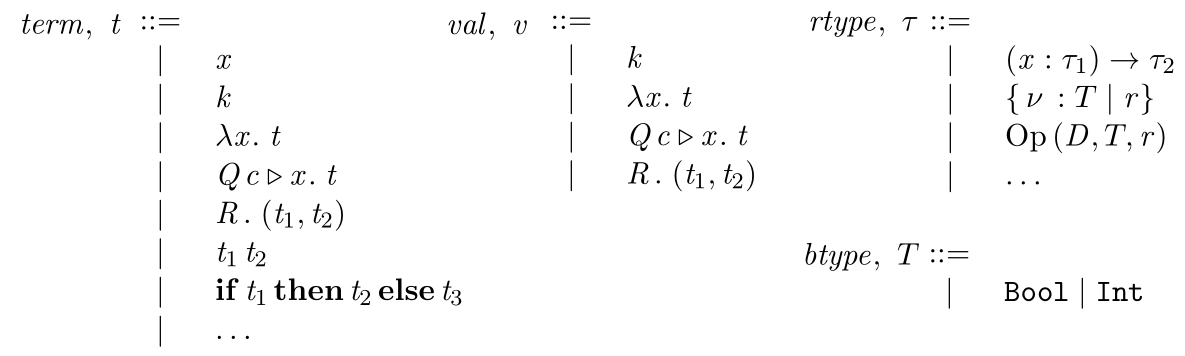}
  \caption[$\lambda^Q syntax$]{Terms, values, and dependent types of
    $\lambda^Q$.  The rules for deriving $\tau$ types for terms are
    found in Figure~\ref{fig:LQ:typing-rules}.  The $k$ metavariable
    represents \texttt{Bool} and \texttt{Int} constants, and $c$
    represents consistency guards.}
  \label{fig:LQ:syntax}
\end{figure}

The $\lambda^Q$ syntax includes two special value terms that interact with a
replicated store.
\begin{description}
\item[Query] The $Q c \triangleright x. t$ term defines an operation
  that queries the global store value up to the consistency predicate
  $c$, binding the value to $x$ before executing the sub-operation
  $t$.  As stated before, the global store value is not explicitly
  stored.  Intuitively, to execute the query, a replica coordinates
  with other replicas ensuring that any effects that violate $c$ are
  either arbitered before the current operation, or after the current
  operation has finished executing.
\item[Return+Emit] The $R. (t_e,t_a)$ term defines a trivial operation which
  performs no query and applies $(t_e,t_a)$ as the operational result,
  in which $t_e$ that is an effect emitted onto the store and $t_a$ is
  a return value that is evaluated and returned to the caller.
  If the $R$ term is nested inside a $Q$ term, the effect and return
  values may include information read from the store.
\end{description}

\begin{example}
  \label{ex:lq:qrterm}
  The basic withdraw bank account operation is expressed in
  $\lambda^Q$ as
  follows:\[ \mathtt{withdraw} \defn \lambda n .\; Q (s_g \geq s_r)
    \triangleright x .\; \mathtt{if}~(x > n)~\mathtt{then}~R
    (\mathtt{Sub}\; n, n)~\mathtt{else}~R (\mathtt{Add}\; 0, 0) \]
  Here, the global store value is queried up to the predicate
  $s_g \geq s_r$, i.e., the value bound to $x$ is at most the global
  value, and the operation is executed assuming that the store value
  is $x$.

  The more involved ``strong'' withdraw operation would be expressed
  as:
  \begin{align*}
    \mathtt{swithdraw} \defn  \lambda n .\; Q (s_g \geq s_r)
    \triangleright x .\; \mathtt{if}~(x >
    n)~&\mathtt{then}~R (\mathtt{Sub}\; n, n) \\[-1ex]
    & \mathtt{else}~Q (s_g = s_r) \triangleright x .\; \mathtt{if}~(x
    > n)&\mathtt{then}~R (\mathtt{Sub}\; n, n)\\[-1ex]
    &   &\mathtt{else}~R (\mathtt{Add}\; 0, 0)
  \end{align*}
  The first query and the \texttt{then} branch act as the standard
  withdraw operation, while the second query (with the stronger
  consistency predicate $s_g = s_r$) learns the exact value of the
  global store (forcing pending deposit operations to commit), and then
  executes the withdraw.
  This operation avoids the stronger coordination needed for the
  second, ``full'' query if it can work safely from just the first
  partial one, while still always making the withdrawal if it's
  absolutely possible.

  For completeness, the deposit operation (which does not need a
  query) would be expressed as
  $\mathtt{deposit} \defn \lambda n .\; R .\; (\mathtt{Add}\; n, n)$.
\end{example}

\begin{figure}[h]
\begin{ottdefnblock}[]{$\Gamma  \vdash  \ottnt{t}  \ottsym{:}  \tau$}{}
\ottusedrule{\ottdruletypeXXvar{}}
  \ottusedrule{\ottdruletypeXXlambda{}} \\[2mm]
  \ottusedrule{\ottdruletypeXXite{}} \\[2mm]
  \ottusedrule{\ottdruletypeXXq{}} \\[2mm]
  \ottusedrule{\ottdruletypeXXr{}} \\[2mm]
\end{ottdefnblock}
  \begin{mathpar}
    \inferrule*[Right=LT-Sub]{
      \Gamma \vdash t : S_1 \\
      \Gamma \vdash S_1 <: S_2 \\
      \Gamma \vdash S_2
    }{
      \Gamma \vdash t : S_2
    }
    \and
    \inferrule*[Right=Dec-$<:$-Base]{
      \text{Valid}(\llbracket \Gamma \rrbracket \land \llbracket t_1
      \rrbracket \Rightarrow \llbracket t_2 \rrbracket)
    }{
      \Gamma \vdash \{\nu : B \;|\; t_1\} <: \{\nu : B \;|\; t_2\}
    }
  \end{mathpar}
  \caption{Typing and sub-typing rules for $\lambda^Q$.}
  \label{fig:LQ:typing-rules}
  \label{fig:lt-sub}
  \vspace{-3ex}
\end{figure}

\subsection{Operation Types}

The type system for $\lambda^Q$ (detailed in
Figure~\ref{fig:LQ:typing-rules}) extends Liquid
Types~\cite{liquidtypes} on the CBV $\lambda$-calculus.
For those unfamiliar, liquid types refine standard types with
predicates on the values.
For example, the typing judgement $t : \{ \nu : \mathtt{Int} \mid x > 5
\}$ asserts that the term $t$ is an integer, as well as that the value
is greater than $5$.

In Figure~\ref{fig:LQ:typing-rules}, standard terms in the language
are typed as per standard liquid types, while CARD operations are
typed under a special $Op$ type.
The typing judgement $t : Op(D, A, \varphi)$ indicates that $t$ is an
operation for the CARD $D$ that returns a value of type $A$
and that any $D$-execution event that results from the operation
satisfies the event specification $\varphi$.

Intuitively, the \textsc{type_q} rule is similar to a conditional
guard rule: if a term $t$ is of type $Op(D,A,\varphi)$ given the
additional premise $\denote{c}$, the term $Q c \triangleright x. t$ is
of type $Op(D, A, \varphi)$.
The \textsc{type_r} rule derives our \texttt{Op} type for a base $R$
term from a standard Liquid Type judgment, stating that the return
value and the denotation of the effect in the $R$ term must together
(in the logical constraint context of $\Gamma$) ensure the \texttt{Op}
type's $\varphi$ specification holds.
The refinement part of this Liquid Type judgment becomes a simple
logical constraint problem according to the rules in
Figure~\ref{fig:lt-sub}.
In these rules, $<:$ is the ``subtype'' relation, which states that
the left hand side has the same basic type as the right hand side, and
that the left's refinement implies the right's refinement.
The denotational brackets on $\llbracket \Gamma \rrbracket$ reduce the
context to the set of logical statements contained in its refinements.

\begin{figure}
  \centering
  \[
    \inferrule*[Right=type_lambda]{
      \inferrule*[Right=type_q]{
        \inferrule*[Right=type_ite]{
          \inferrule{
          }{
            n:\mathtt{Nat}, x:\{\nu:S(\mathtt{Counter})\;|\;x \le s\}
            \vdash (x \ge n) : \mathtt{Bool}
          } \\
          n:\mathtt{Nat}, x:\{\nu:S(\mathtt{Counter})\;|\;x \le s\},x
          \ge n \vdash \{\ldots(\emph{then})\} \\
          n:\mathtt{Nat}, x:\{\nu:S(\mathtt{Counter})\;|\;x \le s\},\neg(x
          \ge n) \vdash \{\ldots(\emph{else})\}
        }{
          n:\mathtt{Nat}, x:\{\nu:S(\mathtt{Counter})\;|\;x \le s\} \vdash \{\mathtt{if}\ldots\} : \mathtt{Op}(\mathtt{Counter},\mathtt{Int},\varphi)
        } \\
        \inferrule{
        }{
          n:\mathtt{Nat} \vdash \mathtt{LE} : C(\mathtt{Counter})
        }
      }{
        n : \mathtt{Nat} \vdash Q\;\mathtt{LE} \triangleright x.\; \{\mathtt{if}\ldots\} \; : \; \mathtt{Op}(\mathtt{Counter},\mathtt{Int},\varphi)
      }
    }{
      \bullet \vdash \lambda n .\; Q\;\mathtt{LE} \triangleright x.\; \{\mathtt{if}\ldots\} \; : \; (n:\mathtt{Nat}) \to
      \mathtt{Op}(\mathtt{Counter},\mathtt{Int},\varphi)
    }
  \]
  \caption{Derivation of \texttt{withdraw} type down to branches with
    base $R$ terms.}
  \label{fig:ex-type-1}
\end{figure}

\begin{figure}
  \centering
  \[
    \inferrule*[Right=type_r]{
      \inferrule{
        \inferrule*[Right=type_var]{
        }{
          \Gamma^+ \vdash n : \mathtt{Nat}
        }
      }{
        \Gamma^+ \vdash \mathtt{Sub}\;n : E(\mathtt{Counter})
      } \\
      \Gamma^+ \vdash n : \{\nu :
      \mathtt{Int}\;|\;s'=\denote{\mathtt{Sub}\;n}(s) \Rightarrow \varphi\}
    }{
      \Gamma^+ \vdash R.\; (\mathtt{Sub}\;n,n) :
      \mathtt{Op}(\mathtt{Counter},\mathtt{Int},\varphi)
    }
  \]
  \caption{Derivation of $R$ term for \texttt{withdraw}'s success
    branch down to standard Liquid Type.}
  \label{fig:ex-type-2}
\end{figure}

\begin{figure}
  \centering
  \[
    \inferrule*[Right=LT-Sub]{
      \inferrule*[Right=Dec-<:-Base]{
        \text{Valid}(\llbracket \Gamma \rrbracket \land \llbracket \mathtt{Nat}
        \rrbracket \Rightarrow \llbracket \{\nu :
      \mathtt{Int}\;|\;s'=\denote{\mathtt{Sub}\;n}(s) \Rightarrow
      \varphi\} \rrbracket)
      }{
        \Gamma^+ \vdash \mathtt{Nat} <: \{\nu :
      \mathtt{Int}\;|\;s'=\denote{\mathtt{Sub}\;n}(s) \Rightarrow
      \varphi\}
      }
    }{
      \Gamma^+ \vdash n : \{\nu :
      \mathtt{Int}\;|\;s'=\denote{\mathtt{Sub}\;n}(s) \Rightarrow
      \varphi\}
    }
  \]
  \caption{Derivation for one of \texttt{withdraw}'s Liquid Type
    obligations into logical constraint problem.}
  \label{fig:ex-type-3}
\end{figure}

\begin{example}
  \label{ex:lq:types}
  As an end-to-end demonstration, we now type-check the
  \texttt{withdraw} operation according to the specfication we have
  been using, for which
  \[\varphi \defn (s \ge 0 \Rightarrow s' \ge 0) \land (a = s - s')\]
  We first follow the derivation in Figure~\ref{fig:ex-type-1},
  storing in the context the constraint on $s$ (the pre-store value)
  that the query on $\mathtt{LE}$ gives us.
  This produces two unsolved branches, one for the \texttt{then}
  branch of the \texttt{if} term on which we can assume $x \ge n$, and
  one on the \texttt{else} branch where we assume the opposite.
  Like the query constraints, these assumptions are added to the
  context.

  We now elide the trivial \texttt{else} branch and follow the
  \texttt{then} branch, referring to the context so far (including
  $x \ge n$) as $\Gamma^{+}$, in Figure~\ref{fig:ex-type-2}.
  This takes us to the standard Liquid Type obligation
  \[
    \Gamma^+ \vdash n : \{\nu :
    \mathtt{Int}\;|\;s'=\denote{\mathtt{Sub}\;n}(s) \Rightarrow
    \varphi\}
  \]
  which may look strange since $n$ already has the type \texttt{Nat}
  in $\Gamma^+$.
  This is where, in Figure~\ref{fig:ex-type-3}, we use the Liquid Type
  subtyping rules to reduce the obligation to a logical constraint
  problem which we can verify by hand or with an SMT solver, and in
  which we are aided by the $s$ constraint from our guarded query:
  \begin{mathpar}
    \llbracket \Gamma^+ \rrbracket \land \llbracket \mathtt{Nat}
        \rrbracket \Rightarrow \llbracket \{\nu :
      \mathtt{Int}\;|\;s'=\denote{\mathtt{Sub}\;n}(s) \Rightarrow
      \varphi\} \rrbracket = \\
    (n \ge 0) \land (x \le s) \land (x \ge n) \land (s' = s - n)
  \Rightarrow (s \ge 0
  \Rightarrow s' \ge 0) \land (a = s -
  s')
  \end{mathpar}
  Deciding this as valid, we have thus verified that \texttt{withdraw}
  has our desired behavior in a concurrent setting.
\end{example}

\subsection{Operation Executions}

$\lambda^Q$ follows the standard semantics of the CBV
$\lambda$-calculus for evaluating standard terms (terms with standard
refinement types, excluding the Op type).
We use the judgement $t \Downarrow_{\lambda} t'$ to represent the
standard big-step semantics for CBV $\lambda$-calculus.

Operations, i.e., terms of type $\text{Op}(D,A,\varphi)$, cannot be
evaluated in a pure setting.
Rather, they are executed by replicas, which may query values from the
global replicated store.
The state of the operational evaluation is represented by
$(s, \psi, t)$ where $s$ is the the global store value, $\psi$ is the
accumulated active consistency guard, and $t$ is the term to be
evaluated.
Each execution step is described abstractly by the \emph{operation
execution rules} (Figure~\ref{fig:LQ:exec-rules}):
\begin{compactitem}
\item Query-evaluation step: A query evaluation step represents a
  replica executing $Q c \triangleright x .\; t$, i.e., querying the
  evaluation global store under the query predicate $c$, and
  evaluating the term $t$ with $x$ bound to the value of the query.
  The replica obtains (non-deterministically, at this level) a value
  $s_x$ such that $c(s_x, s_r)$ holds, and the value of $\psi$ is
  updated with $[s_x / s_r]c$ and the resulting term is obtained by
  substituting the value $s_r$ in $t$.
\item Drift step: A drift step represents the value of the global store
  value changing due to the execution of a different replica.
  However, the $\psi$ value in the execution context restricts the
  change so that snapshots which have been substituted into $t$ (by
  steps of the \textsc{query} rule) remain consistent according to the
  guards they were queried with.
  Note that this rule makes the execution non-deterministic.
\end{compactitem}
Fully executing an operation $t$ with type $\text{Op}(C,A,\varphi)$
from $(s, \top, t)$ produces $(s, \psi, R. (e, a))$ where $a:A$ is the return
value and $\denote{e}(s)$ is the final value of the global store.  By the
soundness of liquid types, we get that
$(s, \denote{e}(s), a) \models \varphi$.

\begin{figure}[h]
  \centering
  \ottdefnprocXXstep{}
  \caption{Operation execution rules}
  \label{fig:LQ:exec-rules}
\end{figure}

\begin{example}
  \label{ex:lq:opexec}
  We describe one execution each of the deposit, withdraw, and strong
  withdraw operations in the bank account example.
  The steps resulting from query and drift steps are
  superscripted with $Q$ and $D$, respectively.
  \begin{compactitem}
  \item The evaluation of $\mathtt{deposit}\; 100$ can produce the
    following sequence:
    $(0, \top, \mathtt{deposit}\; 100) \mapsto^{Q}  (0, \top, R
    (\mathtt{Add}\; 100, -100))
    $
  \item The evaluation of $\mathtt{withdraw}\; 10$ can produce the
    following sequence:
    $(0, \top, \mathtt{withdraw}\; 10) \mapsto^{D} (100, \top,
    \mathtt{withdraw}\; 10) \mapsto^{Q} (100, s_g \geq 100, R
    (\mathtt{Sub}\; 10, 10))$
  \item The nested queries in $\mathtt{swithdraw}$ lead to multiple
    query steps in the  evaluation.
    The following is a valid evaluation sequence:
    $(0, \top, \mathtt{swithdraw}\; 10) \mapsto^{Q} (0, s_g \geq 0,
    t_{iq}) \mapsto^{D} (100, s_g \geq 0, t_{iq}) \mapsto^{D} (90, s_g
    \geq 0, t_{iq}) \mapsto^{D} (90, s_g \geq 0 \land s_g = 90, R
    (\mathtt{Sub}\; 10, 10))$
    where $t_{iq} \defn Q (s_g = s_r) \triangleright x .\;
    \mathtt{if}~(x > 10)~\mathtt{then}~R (\mathtt{Sub}\; 10,
    10)~\mathtt{else}~R (\mathtt{Add}\; 0, 0)$.
  \end{compactitem}
\end{example}

\paragraph{Combining Multiple Operational Executions.}
The operation execution rules produce a sequence of evaluation steps
corresponding to the invocation of a single operation.  We now
describe how a number of different (possibly concurrent) operation
invocations correspond to a CARD execution.  Intuitively, the CARD
execution must be produced by combining the update steps of an
operation execution for each invocation.  The drift steps in the
operation execution of $t$ correspond exactly to the updates of all
the operations arbitrated before the effect produced by $t$, and the
query steps must take as their $s_x$ value a post-store of some subset
of the effects arbitrated before.

Given a set of $D$-operation invocations $T$ with
$\op : T \to \mathtt{Op}(D,A,\varphi)$ giving the operation term for
each invocation, we say a CARD execution $L = (s_0,W,G,\grd,\ar,\vis)$
is \emph{produced by} $T$ iff there exists a one-to-one correspondence
between events $\eta_i \in W$ and operation invocations $t_i \in T$
such that:
\begin{compactitem}
\item there exists an operation execution for $t_i$ of the form
  $(s_0,\top,\op(t_i)) \mapsto^{*} (s_i, \psi_i, R (e_i, a_i))$ in
  which
  $\psi_i=[s_{x0}/s_r]c_0 \land [s_{x1}/s_r]c_1 \land \ldots \land
  [s_{xn}/s_r]c_n$,

\item $\grd(\eta_i)$ contains $n$ active guards corresponding to the
  $n$ clauses in $\psi_i$; $g_j$ corresponds to clause
  $[s_{xj}/s_r]c_j$ in $\psi_i$ such that $\gc(g_j)=c_j$,
  $\eval(L_{g_j})=s_j$ and $\vis^{-1}(g_j)$ includes the $\vis^{-1}$
  set of each guard of each event in $\vis^{-1}(g_j)$.
\item the \textsc{drift} steps in $t_i$'s operation execution
  correspond, in order, to the preceeding events in $L_{\eta_i}$ such
  that for $\eta_j\in L_{\eta_i}$, $\eff(\eta_j)$ is the effect
  quantified in the corresponding \textsc{drift} step's premise,
\item $\eval(L_{\eta_i})=s_i$,
\item $\eff(\eta_i)=e_i$, and
\item $\rval(\eta_i)=a_i$.
\end{compactitem}

\begin{figure}
  \usetikzlibrary{decorations, arrows}

\begin{center}
\hspace*{-4mm}%
\begin{tikzpicture}[font=\tiny]

  \tikzset{cpt/.style={draw=black,fill=white,circle,inner sep=.5mm}}

  \def\shortdrop{3mm}
  \def\longdrop{5mm}
  \def\repgap{36mm}
  \def\repgapp{36.45mm}
  \def\ced{red}
  \def\cew{blue}
  \def\cesw{purple}

  \node[cpt,label=right:{$(0, \top, \mathtt{deposit}\; 100)$}] (r10) [left=-50mm] {};
  \node[cpt,draw=\ced,label={[\ced]right:{$(0, \top, R.(\mathtt{Add}\; 100, -100))$}}] (r1q) [below=\shortdrop of r10] {};

  \node[cpt,label=right:{$(0, \top, \mathtt{withdraw}\; 10)$}] (r20) [right=\repgap of r10] {};
  \node[cpt,draw=\ced,label={[\ced]right:{$(100, \top, \mathtt{withdraw}\; 10)$}}] (r2d) [below right=\longdrop and \repgapp of r1q] {};
  \node[cpt,draw=\cew,label={[\cew]right:{$(100, s_g \geq 100, R.(\mathtt{Sub}\; 10, 10))$}}] (r2q) [below=\shortdrop of r2d] {};

  \node[cpt,label=right:{$(0, \top, \mathtt{swithdraw}\; 10)$}] (r30) [right=\repgap of r20] {};
  \node[cpt,label=right:{$(0, s_g \geq 0, Q\, \guarded{\mathtt{EQ}}{\cdots})$}] (r3q1) [below=\shortdrop of r30] {};
  \node[cpt,draw=\ced,label={[\ced]right:{$(100, s_g \geq 0, Q\, \guarded{\mathtt{EQ}}{\cdots})$}}] (r3d1) [right=\repgap of r2d] {};
  \node[cpt,draw=\cew,label={[\cew]right:{$(90, s_g \geq 0, Q\, \guarded{\mathtt{EQ}}{\cdots})$}}] (r3d2) [below right=\longdrop and \repgapp of r2q] {};
  \node[cpt,draw=\cesw,label={[\cesw]right:{$(90, s_g \geq 0 \land s_g = 90,R.(\mathtt{Sub}\; 10, 10))$}}] (r3q2) [below=\shortdrop of r3d2] {};

  \node[] (r1e) at (r10 |- r3q2) {};
  \node[] (r2e) at (r20 |- r3q2) {};

  \node[cpt,label=right:{$(0, \top, 0)$}] (aex0) [right=(1.2*\repgap) of r30] {};
  \node[cpt,\ced,label={[\ced]right:{$(100, \top, -100)$}}] (aexd) at (aex0 |- r1q) {};
  \node[cpt,\cew,label={[\cew]right:{$(90, s_g \geq 100, 10)$}}] (aexw) at (aex0 |- r2q) {};
  \node[cpt,\cesw,label={[\cesw]right:{$(80, s_g \geq 0 \land s_g = 90, 10)$}}] (aexsw) at (aex0 |- r3q2) {};

  \draw (r10) -- node[left,\ced]{Q} (r1q) -- (r1e);
  \draw (r20) -- node[left,\ced]{D} (r2d) -- node[left,\cew]{Q} (r2q) -- (r2e);
  \draw (r30) -- node[left]{Q} (r3q1) -- node[left,\ced]{D} (r3d1) -- node[left,\cew,yshift=-2mm]{D} (r3d2) -- node[left,\cesw]{Q} (r3q2);
  \draw[dashed] (aex0) -- node[right,\ced]{$\mathbf{\eta_d}$} (aexd) -- node[right,\cew]{$\mathbf{\eta_w}$} (aexw) -- node[right,\cesw]{$\mathbf{\eta_{sw}}$} (aexsw);

  \draw[->,\ced] (r1q) to[out=-25,in=180] node[below]{broadcast $\eta_d$} ++(\repgap/2, -\longdrop);
  \draw[->,\cew] (r2q) to[out=-25,in=180] node[below]{broadcast $\eta_w$} ++(\repgap/2, -\longdrop);

  \node[label=right:{\small r1:\quad$\mathtt{deposit}\; 100$}] (r1) [above=\shortdrop of r10] {};
  \node[label=right:{\small r2:\quad$\mathtt{withdraw}\; 10$}] (r2) [above=\shortdrop of r20] {};
  \node[label=right:{\small r3:\quad$\mathtt{swithdraw}\; 10$}] (r3) [above=\shortdrop of r30] {};
  \node[label=right:{\small abstract execution}] (aex) [above=\shortdrop of aex0] {};
\end{tikzpicture}
\end{center}
  \caption{Correspondence between operational executions of
  $\mathtt{deposit}\; 100$, $\mathtt{withdraw}\; 10$, and
  $\mathtt{swithdraw}\; 10$ and an abstract execution.}
  \label{fig:op-execs-to-card-exec}
\end{figure}
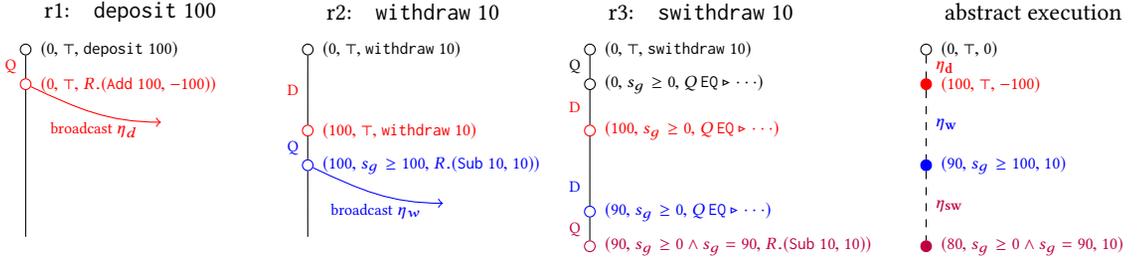

\begin{example}
  \label{ex:lq:op-execs-to-card-exec}
  The operational executions of the deposit, withdraw and strong
  withdraw operations from Example~\ref{ex:lq:opexec} can produce the
  abstract execution
  $L = (0, \{\eta_d, \eta_w, \eta_{sw}\},G,\grd,\ar,\vis)$ where:
  \begin{inparaenum}[(a)]
  \item $\eta_d \defn (\top, \lambda s .\; s + 100, -100)$,
  \item $\eta_w \defn (s_g \geq 100, \lambda s .\; s - 10, 10)$, and
  \item $\eta_{sw} \defn (s_g \geq 0 \land s_g \geq 90, \lambda s .\;
    s - 10, 10)$.
  \end{inparaenum}
  The exact correspondence between the abstract execution and the
  operational executions is depicted in
  Figure~\ref{fig:op-execs-to-card-exec}.
\end{example}

\begin{theorem}[Well-Formedness of Operation Executions]
  \label{thm:op-rules-wf}
  Any $D$-execution $L$ that is produced by a set of $D$-operations
  $T$ is well-formed (by the definition in
  Section~\ref{sec:cards-executions}).
  \proof{
    The non-trivial part is guard-compliance.
    We prove guard-compliance by induction on the operation execution
    step sequence corresponding to each event $\eta$, with
    I.H.  $s \models \psi$.
    \begin{itemize}
    \item Base: $s=s_0$, $\psi$ is empty, trivially satisfied.
    \item Step with \textsc{query}: I.H. gives $s \models \psi$,
      \textsc{query} premise gives $s \models [s_x/s_r]c$, thus $s
      \models \psi \land [s_x/s_r]c$.
    \item Step with \textsc{drift}: Premise gives $s' \models \psi$.
    \end{itemize}
    The pre-store of $\eta$ must be equal to the $s$ value of it's
    operation's final context because each event in $L_{\eta}$ applies
    the same effect as its corresponding \textsc{drift} step.
    The vis-store of any $g\in \grd(\eta)$ is equal to the $s_x$ in
    it's $\psi$ clause by definition of producing a $D$-execution.
    Thus all guards in $L$ are satisfied by their pre-store and
    vis-store values. $\Box$
  }
\end{theorem}

\begin{theorem}[Preservation for Operation Executions]
  \label{thm:soundness2}
  For any derived term $\Gamma \vdash t : \text{Op}(C,A,\varphi)$ and
  starting state $s_0$, if an operation execution
  $(s_0,\top,t) \longmapsto^{*} (s',\psi',R. (e,a))$ exists, then
  $\psi' \Rightarrow \varphi(s',\denote{e}(s'),a)$.
  \proof{
    We must show that $\psi$ is made strong enough to guarantee
    $\varphi$ for a term $\Gamma \vdash t : \text{Op}(D,A,\varphi)$.
    We begin by inductively evaluating and analyzing the type
    derivation of $t$ side by side, showing that at each step,
    $\denote{\Gamma} \Rightarrow \psi$.
    \begin{description}
    \item[Base:] $\Gamma=\psi=\top$.
    \item[Case $Q c \triangleright x.\; t'$:] We evaluate this term by
      a \textsc{query} step, adding $[s_x/s_r]c$ to $\psi$ and
      replacing $x$ with $s_x$ in $t'$.
      We type this term by the \textsc{type_q} rule, adding $[x/s_r]c$
      to $\denote{\Gamma}$.
      So our knowledge of $s_x$ in the evaluated $t'$ is matched by
      our knowledge of $x$ in the typed $t'$, and $(\denote{\Gamma}
      \Rightarrow \psi) \Rightarrow (\denote{\Gamma}\land [x/s_r]c
      \Rightarrow \psi \land [s_x/s_r]c$.
    \item[Case (any other):] This term is evaluated by the standard
      $\lambda$-calculus rules and does not add any obligations to
      $\psi$.
    \end{description}
    We have thus evaluated $t$ to a configuration
    $(s,\psi,R.\;(t_e,t_a))$ and followed its type derivation to a
    term $\Gamma \vdash R.\;(t_e,t_a) : \text{Op}(D,A,\varphi)$ such
    that $\denote{\Gamma} \Rightarrow \psi$ (when $x$'s in $\Gamma$
    are replaced with their corresponding $s_x$'s).
    The remaining obligation of the type derivation shows that the
    contents of $\denote{\Gamma}$ ensure that the final term satisfies
    $\varphi$ under any compatible store value, and so $\psi$ must be
    strong enough to ensure the same
    (Def.~\ref{def:event-sat}). $\Box$
  }
\end{theorem}

\begin{theorem}[Produced $D$-Events Satisfy Operation Specifications]
  \label{thm:prod-event-sat}
  Given an operation invocation $t_i$ in a set of invocations $T$ for
  which $\op(t_i):\texttt{Op}(D,A,\varphi)$, the event $\eta_i$
  corresponding to $t_i$ in any $D$-execution $L$ produced by $T$ via
  the operation execution rules satisfies $\varphi$ (in the sense of
  Def.~\ref{def:event-sat}).
  \proof{
    By Theorem~\ref{thm:soundness2}, we know that for any operation
    execution step sequence for $t_i$ ending with
    $(s',\psi',R.\;(t_e,t_a))$, we have
    $\psi' \Rightarrow \varphi(s',\denote{t_e}(s'),t_a)$.
    And so have this statement for the operation execution sequence
    that produces $\eta_i$, for which $\eff(\eta_i)=t_e$ and
    $\rval(\eta_i)=t_a$.
    The guards in $\grd(\eta_i)$ are together satisfied by the same
    store values that $\psi'$ is satisfied by, and so guard compliance
    (a component of well-formedness of $L$, which we have by
    Theorem~\ref{thm:op-rules-wf}) ensures that
    $\psi'(\eval(L_{\eta_i}]))$.
    Thus for $s=\eval(L_{\eta_i})$ we have
    $\varphi(s,\denote{\eff(\eta_i)}(s),\rval(a))$, meaning that
    $\eta \models_L \varphi$. $\Box$
  }
\end{theorem}

Because operation-produced events respect to their specifications, it
is easy to show that invariants can be maintained.

\begin{theorem}[Execution Invariants]
  \label{lem:invariants}
  Given a $D$-store predicate $I$ and a set of $D$-operation
  invocations $T$, each of which has a type which includes
  $I(s) \Rightarrow I(s')$ in its specification, any $D$-execution, which is
  produced by $T$ and for which $I(s_0)$ holds, preserves $I$.
  \proof{
    This follows immediately from Theorem~\ref{thm:prod-event-sat}.
    Every event in the produced execution will respect
    $I(s) \Rightarrow I(s')$, and so $I$ is preserved over each effect
    application.
  }
\end{theorem}

\begin{example}
  \label{ex:invariants}
  Suppose we want to ensure that the invariant $I \defn s \geq 0$ holds for the
  bank account example, i.e., that the account value is always non-negative. The
  key insight from Theorem~\ref{lem:invariants} is that the task of ensuring
  this invariant can be split into guaranteeing two separate properties:
  \begin{compactitem}
  \item the system only produces events that are
    sound for the specification $I(s) \implies I(s')$, and
  \item the executions are well-formed.
  \end{compactitem}
  For example, if every event produced by the system is in one of the forms of
  $\eta_w$ or $\eta_d$ from Section~\ref{sec:cards} (with the constants $10$ and
  $100$ replaced by any non-negative integer), all these events are guaranteed
  to be sound for the specification. Further, the system would need to ensure
  that these events are executed only in the contexts where the guards hold.
  \end{example}


\sectionnewpage
\section{Inferring Conflict Avoidance Requirements}
\label{sec:locks}

The specifications verified for operations in Section~\ref{sec:lang}
depend on query guards being maintained while concurrent events enter
the execution history.
It is simple to state this requirement in the operation execution
rules, in which each new event is appended in order to the evolving
store value, but we need a more complete picture of effect-guard
interactions in order to design a realistic system in which events
will appear to replicas out of order.

\subsection{Measures of Non-Conflict}

First, we define the following notion of an \emph{immediate accord} between an
effect and a guard.
An immediate accord existing between an effect $e:E$ and a guard $c:C$ implies
that the effect updating the global store cannot violate the consistency guard
in an execution of an action bound by a $c$ query, i.e., actions of the form $Q c
\triangleright x . t$.

\begin{definition}[Immediate Accord]
  Given a CARD $D=(S,E,C)$, guard $c:C$ and effect $e:E$, an
  \emph{immediate accord} exists between them, written as
  $\text{IA}(c,e)$, iff
  \[
    \forall s_g, s_r : S. \; c(s_g,s_r) \implies
    c(\denote{e}(s_g),s_r).
  \]
\end{definition}

We denote by $\text{IAS}_D(c)$ the set of all $D$-effects in immediate
accord with $c$.

\begin{example}
  \label{ex:locks:ia}
  In the running example, there is an immediate accord between the
  effect $\mathtt{Add}\; n$ and the guard $\mathtt{LE}$.  However,
  there is no immediate accord between $\mathtt{Add}\; n$ and
  $\mathtt{EQ}$, or between $\mathtt{Sub}\; n$ and either of
  $\mathtt{EQ}$ and $\mathtt{LE}$.
\end{example}

\begin{definition}[Careful Executions]
  We call a $D$-execution $D=(L,G,\grd,\ar,\vis)$ \emph{careful} iff
  for each $g \in G$ guarding an event $\eta$, $L_g$ contains all
  events $\eta_i$ in $L_{\eta}$ for which $\eff(\eta_i)$ is not in
  immediate accord with $\gc(g)$.
\end{definition}

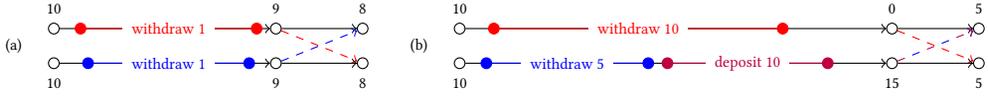
\begin{figure}
  { 

\def\repgap{3mm}
\def\oplen{2.8cm}
\def\synclen{1cm}

\begin{tabular}{l l}

  \begin{tikzpicture}[font=\tiny]
    \tikzset{cpt/.style={draw=black,fill=white,circle,inner sep=.5mm}}

    \node[cpt,label={10}] (r1s) {};
    \node[cpt,label={9}] (r1o) [right=\oplen of r1s]{};
    \node[cpt,label={8}] (r1e) [right=\synclen of r1o]{};

    \node[cpt,label=below:{10}] (r2s) [below=\repgap of r1s] {};
    \node[cpt,label=below:{9}] (r2o) [below=\repgap of r1o] {};
    \node[cpt,label=below:{8}] (r2e) [below=\repgap of r1e] {};

    \draw[->] (r1s) -- (r1o);
    \draw[->] (r1o) -- (r1e);
    \draw[dashed,->,red] (r1o) -- (r2e);

    \draw[->] (r2s) -- (r2o);
    \draw[->] (r2o) -- (r2e);
    \draw[dashed,->,blue] (r2o) -- (r1e);

    \node[cpt,red] (r1oa) [right=2mm of r1s] {};
    \node[cpt,red] (r1ob) [left=1mm of r1o] {};
    \draw[red] (r1oa) -- node[fill=white] {withdraw 1} (r1ob);

    \node[cpt,blue] (r2oa) [right=3mm of r2s] {};
    \node[cpt,blue] (r2ob) [left=2mm of r2o] {};
    \draw[blue] (r2oa) -- node[fill=white] {withdraw 1} (r2ob);

    \path (r1s) -- node[left=\repgap] {(a)} (r2s);
  \end{tikzpicture} & %
  \begin{tikzpicture}[font=\tiny]
    \tikzset{cpt/.style={draw=black,fill=white,circle,inner sep=.5mm}}

    \node[cpt,label={10}] (r1s) {};
    \node[cpt,label={0}] (r1o) [right=(2*\oplen) of r1s]{};
    \node[cpt,label={5}] (r1e) [right=\synclen of r1o]{};

    \node[cpt,label=below:{10}] (r2s) [below=\repgap of r1s] {};
    \node[cpt,label=below:{15}] (r2o) [below=\repgap of r1o] {};
    \node[cpt,label=below:{5}] (r2e) [below=\repgap of r1e] {};

    \draw[->] (r1s) -- (r1o);
    \draw[->] (r1o) -- (r1e);
    \draw[red,dashed,->] (r1o) -- (r2e);

    \draw[->] (r2s) -- (r2o);
    \draw[->] (r2o) -- (r2e);
    \draw[blue,dash pattern=on 3pt off 9pt,dash phase=6pt,->] (r2o) -- (r1e);
    \draw[purple,dash pattern=on 3pt off 9pt,->] (r2o) -- (r1e);

    \node[cpt,red] (r1oa) [right=3mm of r1s] {};
    \node[cpt,red] (r1ob) [left=13mm of r1o] {};
    \draw[red] (r1oa) -- node[fill=white] {withdraw 10} (r1ob);

    \node[cpt,blue] (r2o1a) [right=2mm of r2s] {};
    \node[cpt,blue] (r2o1b) [right=20mm of r2o1a] {};
    \draw[blue] (r2o1a) -- node[fill=white] {withdraw 5} (r2o1b);

    \node[cpt,purple] (r2o2a) [right=1mm of r2o1b] {};
    \node[cpt,purple] (r2o2b) [left=7mm of r2o] {};
    \draw[purple] (r2o2a) -- node[fill=white] {deposit 10} (r2o2b);

    \path (r1s) -- node[left=\repgap] {(b)} (r2s);
  \end{tikzpicture}

\end{tabular}

}
  \caption{Blind luck executions. There can be executions which have events of
  not-in-accord operations that are invisible to one another, and still
  produce a well-defined result.
  }
  \label{fig:blind-luck-execs}
\end{figure}

A careful execution is always produced when a replica resolving a
query must see every event in the network which is not in immediate
accord with its guard.
This safety measure over-approximates the guard satisfaction condition
followed by the operation rules by excluding invisible subsets that
satisfy the guard ``by blind luck'', such as an invisible
account-emptying withdrawal followed by an invisible deposit that
undoes it (see Figure~\ref{fig:blind-luck-execs}).
Intuitively, allowing an undetected ``lucky pair'' also allows an
undetected ``unlucky single'' which would make the query resolution
unsound.
We thus use the careful, well-formed $D$-execution as our basis for
the following definitions.

\paragraph{Transitive Accords.}
As illustrated in Section~\ref{sec:overview} (the joint account
\CARD), it is not sufficient for a replica maintaining $c$ to
coordinate with replicas concurrently emitting effects
$e \not\in \text{IAS}_D(c)$.  A second effect $e' \in \text{IAS}_D(c)$
that is concurrent to $e$ might change the behavior of $e$ if it is
arbitrated earlier.  Hence, we now describe a stronger notion of
accords.

\begin{definition}[Transitive Accord]
  A \emph{transitive accord} exists between an effect $e:E$ and a
  guard $c:C$ (written as $\text{TA}_D(e,c)$) iff for any careful
  $D$-execution $L=(W,G,\grd,\ar,\vis)$ containing an event $\eta$
  guarded by $g$ with $\gc(g)=c$, and for any event $\eta' \notin W$
  for which $\eff(\eta')=e$, the guard $g$ remains satisfied in
  $L'=(W \cup \{\eta'\},G,\grd,\ar \cup \{(\eta',\eta)\},\vis)$.
\end{definition}

A \emph{transitive accord set} for $c$ is a set of effects for
which transitive accords exist.
Intuitively, any replica maintaining a guard $c$ needs to coordinate
with replicas emitting effects which are not in its transitive accord
set because a new event arriving at the replica may be inserted
somewhere in the middle of history by the arbitrary ordering.
The following theorem states that finding the largest transitive
accord set is undecidable.

\begin{theorem}
  \label{thm:undecidability}
  Given a CARD $D$ and $D$-guard $c$, finding the largest cardinality
  transitive accord set for $c$ is undecidable.
  \proof{Sketch: the proof relies on constructing an effect $e$ which can
  induce a violation of the guard $g$ only from a single store state.
  Now, $e \in \text{TAS}(c)$ if and only if that single store state is
  reachable through the effects of the system.
  Such store value reachability problems are undecidable.
  }
\end{theorem}

\begin{example}
  \label{ex:locks:ta}
  In the joint bank account example, let's intuit the transitive accord set
  for the guard of \texttt{withdrawJ}, $c = \mathtt{LE} \land \mathtt{App?}$.
  Recall that the state is expressed as a tuple $(s: \mathtt{Int}, b_1:
  \mathtt{Bool}, b_2: \mathtt{Bool})$, and that $\mathtt{withdrawJ}
  \defn Q\;\mathtt{LE}\land\mathtt{App?}\triangleright(\cdots)$, where
  $\denote{\mathtt{LE}} \defn s(s_r) \leq s(s_g)$ and $\denote{\mathtt{App?}}
  \defn b_2(s_r) = b_2(s_g)$. We begin by deciding the immediate accord set of
  $c$, $\text{IAS}_D(c)$:
  \begin{itemize}
    \item $\denote{\mathtt{Request}}$ only changes $b_1$, which is not used in
      either of \texttt{withdrawJ}'s guards. Therefore the effect is in
      $\text{IAS}_D(c).$
    \item $\denote{\mathtt{Approve}}$ and $\denote{\mathtt{Reset}}$ can both
      change $b_2$, violating \texttt{App?}, so neither is in $\text{IAS}_D(c)$.
    \item $\denote{\mathtt{Add}\ n}$ only increases $s$, satisfying
      $\mathtt{LE}$ and $\mathtt{App?}$ (trivially), so it is in
      $\text{IAS}_D(c).$
    \item $\denote{\mathtt{Sub}\ n}$ and $\denote{\mathtt{Set}\ n}$ can both
      decrease $s$, violating \texttt{LE}, so neither is in $\text{IAS}_D(c).$
  \end{itemize}
  Therefore, the immediate accord set of $\mathtt{LE} \land \mathtt{App?}$
  contains $\mathtt{Request}$ and $\mathtt{Add}\ n$. Now let's see which of
  these two is also in the transitive accord set. Notice the presence of
  an additional $\mathtt{Add}\ n$ can never decrease $s$, even when combined
  with other rules. Nor can it change $b_2.$ This shows that $\mathtt{TA}_D(
  \mathtt{Add}\ n, c).$ $\mathtt{Request}$ is more complicated, since it
  toggles $b_1$, which sets $b_2$ when combined with $\mathtt{Approve}.$
  Consider an abstract execution consisting of an $\mathtt{Approve}$ followed
  by a $\mathtt{Withdraw}\ 10.$ Because there is no request, $b_1 = \bot$, the
  $\mathtt{Approve}$ will keep $b_2 = \bot.$ This will result in the
  $\mathtt{Withdraw}\ 10$ acting as a $\mathtt{NoOp}$ Now suppose we produce a
  new execution using the same events preceded by a $\mathtt{Request}.$ This
  time $b_1 = \top$, and could lead to the withdraw being executed.
  Therefore, the only effect with a transitive accord with $\mathtt{LE} \land
  \mathtt{App?}$ is $\mathtt{Add}.$
\end{example}

\subsection{Inferring Minimal Locking Conditions}
\label{sec:locks-minimal}

\paragraph{Consistency Invariants}
A \emph{consistency invariant} in a CARD $D$ is a $D$-guard $c$ for
which, given any pair of $D$-states $(s_g,s_r)$ and $D$-effect $e$,
$c(s_g,s_r) \Rightarrow c(\denote{e}(s_g),\denote{e}(s_r))$.

\begin{theorem}[CINV + IA = TA]
  For a CARD $D=(S,E,C)$, if a $c:C$ is a consistency invariant in $D$
  and an effect $e:E$ is in immediate accord with $c$, then $e$ is
  also in transitive accord with $c$.
  \proof{
    Suppose we have a careful, well-formed $D$-execution $L$
    containing event $\eta$ and active guard $g \in \grd(\eta)$ for
    which $\gc(g)$ is a consistency invariant.
    As $L$ is well-formed, $g$ is satisfied, meaning that
    $\gc(g)(\eval(L),\eval(L_g))$ holds.

    We now take a new event $\eta_2$ for which
    $\text{IA}(\gc(g),\eff(\eta_m))$ holds and create a new execution
    $M=(W \cup \{\eta_m\},G,\grd,\ar \cup \{\eta_m \times
    \eta\},\vis)$.
    Showing that $g$ is also satisfied in $M$ is proof that
    $\text{TA}(\gc(g),\eff(\eta_m))$ holds.
    We show this by inductively evaluating $M_{\eta}$ and $L_g$
    alongside each other and noting that at each step, the post-states
    of the two sub-executions satisfy $g$'s consistency guard.
    This will give us that $\gc(g)(\eval(M_{\eta}),\eval(L_{g}))$,
    showing that $g$ is satisfied in $M$.

    At the base case, $\gc(g)(s_0,s_0)$ holds by definition of
    consistency guards (they are always implied by equality).
    For our inductive step, we examine an event $\eta'$ which is in
    some combination of the executions $L_{\eta}$, $M_{\eta}$, and
    $L_{g}$, with $\gc(g)(s_M,s_{L_g})$ as our inductive hypothesis:
    \begin{description}
    \item[Case $\eta' \in L_{\eta} \cap M_{\eta} \cap L_{g}$:] The
      fact that $\gc(g)$ is a consistency invariant gives us\\
      $\gc(g)(\denote{\eff(\eta')}(s_M),\denote{\eff(\eta')}(s_{L_g}))$.
    \item[Case
      $\eta' \in L_{\eta} \cap M_{\eta} \land \eta' \notin L_{g}$:]
      Because $L$ is careful and $\eta'$ is not in $L_{g}$, we must
      have $\text{IA}(\gc(g),\eff(\eta'))$.  This gives us
      $\gc(g)(\denote{\eff(\eta')}(s_M),s_{L_g})$.
    \item[Case
      $\eta' \in M_{\eta} \land \eta' \notin L_{\eta} \cup L_{xi}$:]
      This can only be our new event $\eta_m$ for which we have
      $\text{IA}(\gc(g),\eff(\eta_m))$ by assumption.  This gives us
      that $c(\denote{\eff(\eta')}(s_M),s_{L_g})$.
    \end{description}
    Thus we have $\text{TA}(\gc(g),\eff(\eta_m))$ because $g$ remains
    satisfied when $\eta_m$ is added to $L$. $\Box$
  }
\end{theorem}

Consistency invariants for CARDs play the role equivalent to standard
inductive loop invariants in sequential program verification --- they
are a strengthening of the required property that is preserved by
operations.  We show that every consistency invariant that implies a
given $c$ defines a transitive accord set for $c$.

\begin{theorem}
  \label{thm:ci-tas}
  Let $D$ be a CARD and $c$ and $c'$ be $D$-guards.  If $c'$ is a
  consistency invariant and $c' \Rightarrow c$, then
  $\text{IAS}_D(c')$ is a transitive accord set for $c$.
\end{theorem}

Note that the identity relation itself ($=$) is always a
consistency invariant, similar to how $\bot$ is always a loop
invariant in the sequential setting.  However, this consistency
invariant leads to a transitive accord set that rejects all state
mutating effects in the \CARD.  The challenge is to identify the
consistency invariant that leads to the most complete transitive
accord set.

In spite of Theorem~\ref{thm:undecidability}, we present a simple
semi-procedure that computes a reasonable transitive accord set
in practice through consistency invariants.  First, let the
\emph{weakest consistency precondition} of a guard $c$ and effect
$e$, $\WCP(e,c)$, be the weakest guard such that
$(s_g,s_r) \models \WCP(e,c)$ implies that
$(\denote{e}(s_g),\denote{e}(s_r)) \models c$.  Now, we decide
transitive accords with:
\[
  \text{TAS}_D(c) \defn \mathbf{let} \; c' = \bigwedge_{e:E} \WCP(e,c) \;
  \mathbf{in} \;
  \mathbf{if} \; c \Rightarrow c' \;
  \mathbf{then} \; \text{IAS}_D(c) \;
  \mathbf{else} \; \text{TAS}_D(c \land c')
\]

The following theorem states the soundness of the above procedure.
\begin{theorem}
  \label{thm:tas-soundness}
  Given a CARD $D$, a $D$-guard $c$, and a $D$-effect $e$, the
  procedure $\text{TAS}_D(e,c)$ returns a transitive accord set for
  $c$.
  \proof{
    The proof follows from the following:
    \begin{itemize}
    \item The guard argument at recursive call $i$ (which we will
      call $c_i$) is a strengthening of $c$.
    \item If, at recursive call $i$, the condition
      $c_i \Rightarrow c'$ holds, then $c_i$ is a consistency
      invariant in $D$ because $\forall e:E.\; c_i \Rightarrow
      \WCP(e,c)$.
    \item Therefore, because $c_i \Rightarrow c$ and $c_i$ is a
      consistency invariant, then the returned $\text{IAS}_D(c_i)$ is
      a transitive accord set for $c$ by Theorem~\ref{thm:ci-tas}.
    \end{itemize}
  }
\end{theorem}

The procedure $\text{TAS}$ is computing the greatest fixed-point
$c'_L$ of the equation
$\mu c' : c' \implies c \land ((s_g, s_r) \models c') \implies
\bigwedge_{e:E} (\denote{e}(s_g), \denote{e}(s_r)) \models c'$ as a
consistency invariant and using it to decide transitive accords.
However, any fixed-point of the equation is sufficient, and any
technique used in standard sequential program reasoning can be applied
to compute this fixed-point (e.g., widening from abstract
interpretation, logical interpolant computation, etc).


\section{Implementing a Replica Network}
\label{sec:replica}

In this section, we show how inferred locking conditions can be used
to implement a network of replicas that correctly execute concurrent
CARD operations.  In Figure~\ref{fig:replica-rules}, we detail the
small-step semantics by which a network of replicas executes
operations, which refines the behavior of the previously defined
operation execution rules.  The semantics leverage the transitive
accord sets computed using the procedure detailed in
Section~\ref{sec:locks}.

\paragraph{Replica-network State.}
We represent the state of a replica in the network as $(r, h_r, ts)$
where:
\begin{inparaenum}[(a)]
\item $r$ is the unique replica id;
\item $h_r$ is the replica's view of the network execution history,
  initially set to the empty history; and
\item $ts$ is the sequence of operations yet to be executed, initialized
  non-deterministically to the set of operations a replica will execute.
\end{inparaenum}
The state of a network is given by $(h\big|\big|ls\big|\big|rs)$ where:
\begin{itemize}
\item The \emph{history} $h$ is an set of events of the form
  $v=\text{event}(r,e,a,h_r)$, in which $r$ is a unique replica ID,
  $e$ is an effect, $a$ is a return value for the operation, and
  $h_r \subseteq (h \setminus \{v\})$ is the part of the history that influenced
  the creation of the event $v$.  The values $h_r$ together represent a DAG of events 
  ordered by \emph{happens-before}.
  Delivering an event $v$ to a replica requires that the replica
  already has all events it depended upon, such that causal
  consistency is maintained.
  Note that the history $h$ is not explicitly stored in any replica,
  and cannot be directly read.
\item The \emph{locking configuration} $\text{ls}$ is a map of replica
  ID $r$ to guard $c$, which describes the network constraints which
  need to be maintained in order to preserve the assumptions of
  operations currently under execution.
  $\text{permits}(\text{ls},e)$ determines whether an effect $e$ can
  possibly invalidate any $c$ in $\text{ls}$, and precisely states
  that $e$ is in transitive accord (see Section~\ref{sec:locks}) with
  all guards in $\text{ls}$ except the emitting replica's.
  Possible implementations of a decision procedure for
  $\text{permits}(\text{ls},e)$ are described in
  Section~\ref{sec:locks}.
\item The \emph{replica set} $\text{rs}$ is the set of replicas in the
  network.  Each replica has a unique ID $r$, its own partial view of
  history $h_r \subseteq h$, and a sequence of operations to execute
  $\text{ts}$.
\end{itemize}

The explicit replica execution rules are shown in
Figure~\ref{fig:replica-rules}.
\begin{compactitem}
\item \emph{Lock acquisition.}
  The \textsc{R_Lock} rule describes the precise condition for a lock
  acquisition.
  The rule adds the guard $c_2$ to the replica's  guards in the lock
  state, in the scenario that there are no events that are present in
  the network, but not in the replica history whose effects are not in
  transitive accord with $c_2$.
  In practice, implementing this rule involves communicating with each
  replica in the network, gathering any effects not in
  $\text{TAS}_D(c_2)$, and acquiring a license from each of them.
\item \emph{Operation evaluation.}
  The \textsc{R_Query} rule describes the local execution of the
  operation in the replica.
\item \emph{Effect emission.}
  In case the lock state permits the emission of the effect, the effect
  emission rule adds an event to the local history.
  A lock state permits an emission of effect $e$ if no replica has a
  lock on a guard $c$ such that $e \not\in \text{TAS}_D(c)$.
  Once the effect is emitted, the lock state is updated by removing all
  locked guards for the replica $r$.
\item \emph{Effect delivery.}
  The effect delivery rule transmits an effect that is in the network
  history into the local history of a replica.
\end{compactitem}
\begin{figure}
\begin{mathpar}
  \inferrule*[Right=r_lock]{
    h \setminus h_r \subseteq \text{TAS}_D(c_2)
  }{
    (h\big|\big|\text{ls},r:c_1\big|\big|\text{rs},(r,h_r,Q \; c_2
    \triangleright x.t :: \text{ts}))
    \longmapsto
    (h\big|\big|\text{ls},r:c_1 \land c_2\big|\big|\text{rs},(r,h_r,Q \; c_2
    \triangleright x.t :: \text{ts}))
  }
  \and
  \inferrule*[Right=r_query]{
    c_1 \Rightarrow c_2 \and
    [\text{eval} \; h_r / x]t \Downarrow_{\lambda} t'
  }{
    (h\big|\big|\text{ls},r:c_1\big|\big|\text{rs},(r,h_r,Q \; c_2
    \triangleright x.t :: \text{ts}))
    \longmapsto
    (h\big|\big|\text{ls},r:c_1\big|\big|\text{rs},(r,h_r,t'
    :: \text{ts}))
  }
  \and
  \inferrule*[Right=r_emit]{
    \text{permits}(\text{ls},e) \and
    v = \text{event}(r,e,a,h_r)
  }{
    (h\big|\big|\text{ls}\big|\big|\text{rs},(r,h_r,R. (e,a)::\text{ts}))
    \longmapsto
    (h + v\big|\big|\text{ls} \setminus r\big|\big|\text{rs},(r,h_r + v,\text{ts}))
  }
  \and
  \inferrule*[Right=r_deliver]{
    v = \text{event}(r_v,e,a,h_v) \and
    h_v \subseteq h_r
  }{
    (h + v\big|\big|\text{ls}\big|\big|\text{rs},(r,n,h_r,\text{ts}))
    \longmapsto
    (h + v\big|\big|\text{ls}\big|\big|\text{rs},(r,n,h_r + v,\text{ts}))
  }
\end{mathpar}
  \caption{Replica execution rules}
  \label{fig:replica-rules}
  \vspace{-2em}
\end{figure}

\paragraph{Locking protocol.}
The replica rules we present here are declarative; they specify
\emph{when} a replica is allowed to proceed with locking or querying
but do not give instructions for actively getting to that state.
For this purpose we can use any distributed locking protocol.
A simple locking scheme would require a replica making a query to
contact all other replicas, requesting from them an agreement to not
emit effects that could violate the querying replica's guard, and
further to immediately send all already-emitted effects that could
violate it.
Upon finishing its operation, the querying replica contacts the others
again to release the agreement and deliver its newly emitted effect,
so that no other node can later emit one that arbitrates before it.
This scheme allows for ``asymmetric'' conflicts, in which only one
type of effect in a conflicting pair is responsible for coordination,
which may be efficient if one is much rarer than the other.

In the case of deadlock induced by nested, conflicting queries on two
replicas, it is always safe to abort and retry an operation because
it makes no change to the system until the final emit step.

\paragraph{Producing Executions}
Like the operation executions rules, the replica network execution
rules non-deterministically produce CARD executions.
Given a set of replicas $R$ with $D$-operation invocation sequences on
a replica network, the invocations make a partially ordered set
$(T,\leq)$ where $t_1 \leq t_2$ iff $t_1$ occurs before $t_2$ in the
invocation sequence on a single replica and with
$\op : T \to \mathtt{Op}(D,A,\varphi)$ giving the operation term for
each invocation.
Then we say a CARD execution $L=(W,G,\grd,\ar,\vis)$ is \emph{produced
by} $(T,\leq)$ from the replica network execution rules iff there
exists a one-to-one correspondence between events $\eta_i \in W$ and
operation invocations $t_i \in T$ such that:
\begin{itemize}
\item $t_i \leq t_j \Rightarrow \ar(\eta_i,\eta_j)$,
\item there exists a replica execution for $R$ which fully evaluates
  all operation invocations,
\item the \textsc{r_emit} replica execution step for an invocation
  $t_i$ of the form
  \[
    (h\big|\big|\text{ls}\big|\big|\text{rs},(r,h_r,R. (e,a)::\text{ts}))
    \longmapsto
    (h + v\big|\big|\text{ls}\big|\big|\text{rs},(r,h_r + v,\text{ts}))
  \]
  corresponds CARD event $\eta_i$ in that $\eff(\eta_i)=e$ and
  $\rval(\eta_i)=a$,
\item the \textsc{r_emit} step for $t_i$ is preceeded by $n$
  \textsc{r_query} steps of the form
  \[
    (h\big|\big|\text{ls},r:c_1\big|\big|\text{rs},(r,h_r,Q \; c_2
    \triangleright x.t :: \text{ts}))
    \longmapsto
    (h\big|\big|\text{ls},r:c_1\big|\big|\text{rs},(r,h_r,[\text{eval} \; h_r / x]t
    :: \text{ts}))
  \]
  on $t_i$'s replica which correspond to the $n$ guards in
  $\grd(\eta_i)$ such that $\vis^{-1}(g_n)$ contains all events in
  $h_r$ and $\gc(g)=c_2$.
\end{itemize}

\begin{theorem}[Replicas Implement Operation Rules]
  \label{thm:impl-op-rules}
  If a CARD execution $L$ is produced by a partially ordered set of
  operations $(T,\leq)$ via the replica rules, then $L$ is also
  produced by $T$ via the operation rules.
  \proof{
    We generate an operation execution rule sequence for each
    invocation $T$ from our replica rule sequence and show that it is
    a proof that $L$ is produced by $T$.

    The operation rule sequence for $t_i \in T$ is created from the
    replica rule steps leading up to $t_i$'s \textsc{emit_r} step as
    follows:
    \begin{itemize}
    \item For every step in the operation rule sequence, the starting
      and ending $s$ values are the evaluations of the starting and
      ending $h$ values of the replica rule step that generated it.
    \item An \textsc{emit_r} step for a different invocation creates a
      \textsc{drift} step.
      We satisfy the ``$\exists e$'' premise of the \textsc{drift}
      step using the $e$ that is emitted in the corresponding
      \textsc{emit_r} step.
      The $\text{permits}(ls,e)$ premise ensures that
      $e(s) \models \psi$, because every $[s_x/s_r]c$ clause in $\psi$
      is overapproximated by the inclusion of $c$ in $\text{ls}(r)$ of
      the replica rule context.
    \item An \textsc{r_query} step that contributes to the evaluation
      of $t_i$ creates a \textsc{query} step.
      The $(s,s_x) \models c$ premise of \textsc{query} is guaranteed
      because the \textsc{r_query} step was preceeded by a
      \textsc{r_lock} step with an equivalent premise, and intervening
      \textsc{r_emit} steps are prevented from invalidating it.
    \item An \textsc{r_query} step that contributes to another
      evaluation is ignored.
    \item An \textsc{r_lock} step is ignored.
    \item An \textsc{r_deliver} step is ignored.
    \end{itemize}
    We now show that this generated operation execution sequence for
    $t_i$ satisfies the execution production requirements.
    \begin{itemize}
    \item \emph{The final $\psi$ value must have a clause for every
        guard in $\grd(\eta_i)$.}
      Every \textsc{r_query} step that adds a guard to $\grd(\eta_i)$
      also adds a \textsc{query} step adding the necessary clause to
      $\psi$.
    \item \emph{The \textsc{drift} steps must correspond to the events
        preceeding $\eta_i$.}
      Each event preceeding $\eta_i$ came from a \textsc{r_emit} step
      in the replica rule sequence, which generated the necessary
      \textsc{emit} step in the operation rule sequence.
    \item \emph{We need $\eval(L_{\eta_i}=s_i)$, $\eff(\eta_i)=e_i$,
        and $\rval(\eta_i)=a_i$ for the final operation rule context.}
      The correspondance of added \textsc{drift} rules to events in
      $L_{\eta_i}$ give the first.  The $e$ and $a$ in the final
      operation rule context matches the $e$ and $a$ in the final
      replica rule context, which are what $\eta_i$ is created from,
      giving the second and third.
    \end{itemize}
    Thus we generate all the necessary evidence that $L$ is produced
    by $T$ via the operation rules. $\Box$
  }
\end{theorem}

\begin{lemma}[Well-Formed Executions from Replicas]
  Any CARD execution $L$ produced by a set of operation invocations $T$
  through the replica network execution rules is well-formed.
  \proof{For $L$ to be produced by $T$ through the replica network
    rules, it must also be produced by $T$ through the operation
    execution rules and thus must be well-formed by
    Theorem~\ref{thm:op-rules-wf}}. $\Box$
\end{lemma}


\section{Conflict Detection Evaluation}
\label{sec:evaluation}

\begin{figure}[h]
  \scriptsize
  \centering
  \begin{tabular}{| l | r | r | r | c |}
    \hline
    \textbf{Application} & \textbf{Guards} & \textbf{Effect Classes} & \textbf{Time (ms)} & \textbf{Minimal?} \\ \hline
    Bank account             & 4 & 3 &  35 & Yes \\ \hline
    Bank account with reset  & 4 & 4 &  33 & Yes \\ \hline
    Conspiring booleans (2)  & 4 & 3 &  31 & Yes \\ \hline
    Joint bank account       & 6 & 8 &  59 & Yes \\ \hline
    KV bank accounts (10)    & 11 & 9 & 175 & Yes \\ \hline
    State machine (3 states) & 3 & 3 &  46 & Yes \\ \hline
  \end{tabular}
  \vspace{1mm}
  \caption[CA experiment results]{Conflict avoidance set inference}
  \label{fig:results}
\end{figure}
We empirically evaluated whether the core computational task necessary for
implementing CARDs --- inferring transitive accord sets --- is efficient and
complete.
We implemented the TAS
algorithm~\footnote{\url{https://github.com/cuplv/dsv}}, using the Z3 SMT solver~\cite{DeMoura:2008} for
logical reasoning.
We modeled CARD applications of varying complexity, and
computed TA sets for their consistency guards.
Our applications were simple SMT-representable data structures using integers,
booleans, and arrays.
Each application's guards included the empty guard, the total guard (the
identity relation), and interesting non-trivial guards required by
operations or providing useful information.
For all tested applications, our solver found TA sets in less than $175$ms.
Manual examination proved that these conflict avoidance sets are the smallest
possible ones.
We now detail the applications tested.

\paragraph{Bank account}

This is the simplest form of our running example, including
\texttt{deposit} and \texttt{withdraw} operations which each take a
positive amount parameter and produce Add or Sub Counter effects.
The guard necessary for \texttt{withdraw} in order to preserve the
positive account invariant, $\mathsf{LE} := s_r \le s_g$, was found to
conflict only with $\mathtt{Sub}$, thus matching the intuitive
reasoning: ``Withdrawals must not be concurrent''.

\paragraph{Joint bank account}

This example models the joint bank account from
Section~\ref{sec:overview} involving the request/approval sequence.
The inference procedure correctly inferred that the TA set for
$\mathsf{LE} := s_r \le s_g$ should include the \texttt{Sub},
\texttt{Request}, and \texttt{Approve} effect classes.

\paragraph{Bank account with reset}

We extended Counter CARD backing the bank account with a
\texttt{Reset} effect which sets the store value to $0$.
Reset never drops the value below $0$ by itself, and thus an operation
can safely (with respect to the bank account invariant) emit a
\texttt{Reset} without looking at the store.
We note two interesting aspects about this example:
\begin{inparaenum}[(a)]
\item Intuitively, \texttt{Reset}s can execute freely on their own,
  but \texttt{Sub} requires coordination to halt \texttt{Reset}s and
  \texttt{Sub}s.
  Our technique automatically infers this: the TA set for
  $\mathsf{LE}$ contains \texttt{Reset}, but the TA set for the
  trivial guard of a safe \texttt{reset} operation is empty.
  This is unlike other mixed-consistency systems such Quelea and
  RedBlue\cite{ECDS,RedBlue} where conflicts are symmetric.
\item Due to the arbitration total ordering, the non-commutability of
  \texttt{Reset} has no impact on SEC.
\end{inparaenum}

\paragraph{Finite state machine}
We modeled a distributed finite state machine with a CARD where $S$ is
the set of states and $E$ is the set of transition labels.
Though the effects are non-commutable, the arbitration maintains SEC
without any coordination.
Now, suppose that we write an operation that reads the state under the
guard $(s_g = s_c \Leftrightarrow s_r = s_c)$, i.e., if the global
state is some critical state $s_c$, the operation is guaranteed to see
it.
The TA set for this new guard includes not only ``offenders'' ---
those operations leading into and out of $s_c$ --- but also any that
determine whether offenders will take that action.
In our case, we used an FSM with $3$ states and $3$ transition
effects, and found that the TA set included the effect $A$ that led
into the critical state, and one other effect $B$ that led to the
state from which $A$ led to the critical state.
Note that executing a new operation which is interested in the
critical state completely changes the coordination behavior of the
CARD application, without any other operations or invariants needing
to be rewritten.

\paragraph{Key-value bank accounts}

This example models an array of ten indexed bank accounts, supporting
the same effect classes as the regular bank account but an additional
index parameter -- the logical reasoning for this example involved
using the array SMT theory.
We inferred TA sets both for guards that constrained the global-local
values of the individual accounts, and for guards that constrained
the global-local values of the summation of the accounts.
The TA set for the summation $\mathsf{LE}$ guard included the \texttt{Sub}
effects for all bank accounts (indices), while the TA sets for the
individual account $\mathsf{LE}$ guards included only the \texttt{Sub} effects
for that bank account.
This illustrates that CARDs allow operations pertaining to different
parts of the state to run in parallel even when they are not purely
conflict-free.


\section{Related Work}
\label{sec:relwork}

We described how our work builds on CRDTs (Shapiro et al.~\cite{CRDTs} provide a comprehensive overview).
Several frameworks allow both conflict-free, and conflicting
operations~\cite{Gotsman:2016,Balegas:2015,Li:2014,ECDS,RedBlue,bayou}, offering different trade-offs between consistency
and availability. Such mixed-consistency systems are typically built upon key-value databases
that offer tunable transaction isolation~\cite{Bailis:2013,Terry:2013,Lakshman:2010}.

Our work is closest to the work of ~\cite{Gotsman:2016}, which also focuses on
on reasoning about data types with such conflicting operations. The approach of
~\cite{Gotsman:2016} allow the programmer to specify for every pair of
operations whether there is a conflict, using a token based system. In contrast,
our consistency guards are specified for each operation separately, which allows
the developer to reason only about the operation they are currently writing.
Note that while our consistency guards (replica state - global state relations)
are related to the guarantee relations (replica state - replica state relations)
of~\cite{Gotsman:2016}, the most important difference is how these are used.
\cite{Gotsman:2016} use the guarantee relations only in the proof of correctness
of a program (as a manual step). The programmer cannot write these guarantees,
they can only declare conflicts explicitly between each pair of operations. In
contrast, our language lets the programmer specify the guards directly,
leading to modular specifications, from which conflicts can be algorithmically
inferred.

The second closest work is that of \cite{Balegas:2015}, introduces explicit
consistency, in which concurrent executions are restricted using an application
invariant. Two technically most important differences are: first, our
consistency guards are significantly more expressive than invariants. The
consistency guards relate the global state to the local state, whereas
invariants talk only about one state. That means that in the framework of
Balegas et al., one cannot specify a property such as ``if {\tt getBalance}
returns a value $v$, then the account balance is at least $v$'' (see the bank
account with interest in Section~\ref{sec:bank-with-interest}). Second, our
consistency predicates allow finding conflicts by checking conditions on sequential programs. In contrast, application invariants of Balegas et
al. require to check conditions on concurrent programs, a significantly harder
task.

A related approach~\cite{ECDS,RedBlue} allows manual selection
of consistency levels for operations.
Quelea~\cite{ECDS} allows specifying contracts (ordering constraints) on
effects. In contrast, our system hides the concept of effect ordering in
history, and allows modular conflict specification.
CARDs can use such systems as a backend,
automatically generating the contracts via the conflict inference technique.

The homeostasis protocol~\cite{homeostasis} addresses conflicts between
operations by allowing bounded inconsistencies as long as other forms of
correctness are preserved. It may be possible to fruitfully combine consistency
guards with relaxed consistency notions. We leave this for future work.

Bayou~\cite{bayou} is an early system for detecting and managing conflicts. The
conflicts are detected (translated to our terminology) by re-running a check on
every replica where an effect is propagated to see if the data has been updated
in parallel. This approach to conflict detection is very different from our
consistency guard (which are predicates that link a global and local state).

The axiomatic specification which we used to define \CARDs is based on the
model presented in~\cite{Burckhardt:2014,Attiya:2016}. We built on the model to
define consistency guard compliance, as well as type checking soundness.
The tension between consistency and availability in distributed systems is captured
by the CAP theorem~\cite{B00,GL12} --- we aim to preserve eventual
consistency, while maximizing availability.

\section{Conclusion}
\label{sec:coclusion}

We present CARDs, a new extension of CRDTs which allow conflicting operations.
The key idea was to develop a language that gives programmers the ability to
specify consistency guards that establish what a CARD operation expects from its
distributed environments. This enables modular and sequential reasoning about
CARD operations.

This paper opens several possible directions for future work. Among these, we
plan to pursue extending our language to allow composition of CARDs, as well as
transactions with multiple emits. We also plan to work on quantitative relaxations of our invariant requirements.
Furthermore, we will investigate systems
aspects of our approach: we will empirically investigate different approaches to implementation of our conflict avoidance algorithm.

\newpage

\ifacm
\bibliographystyle{ACM-Reference-Format}
\else
\bibliographystyle{plain}
\fi
\bibliography{shared}

\end{document}